\documentclass[aps,prd,nofootinbib,superscriptaddress]{revtex4}

\usepackage{amsfonts,amssymb,amsthm,bbm,amsmath}
\usepackage{color,psfrag}
\usepackage[dvips]{graphicx}

\newcommand{\C}{{\mathbb C}}
\newcommand{\N}{{\mathbb N}}
\newcommand{\R}{{\mathbb R}}

\newcommand{\cA}{{\mathcal A}}

\newcommand{\cV}{{\mathcal V}}

\newcommand{\cH}{{\mathcal H}}

\newcommand{\cS}{{\mathcal S}}
\newcommand{\SU}{\mathrm{SU}}

\newcommand{\SL}{\mathrm{SL}}

\newcommand{\U}{\mathrm{U}}

\newcommand{\be}{\begin{equation}}
\newcommand{\ee}{\end{equation}}
\newcommand{\beq}{\begin{eqnarray}}
\newcommand{\eeq}{\end{eqnarray}}
\newcommand{\bea}{\begin{eqnarray}}
\newcommand{\eea}{\end{eqnarray}}
\newcommand{\nn}{\nonumber}

\newcommand{\mat} [2] {\left ( \begin{array}{#1}#2\end{array} \right ) }

\newcommand{\bin} [2] {\left (\begin{array}{c}#2\\#1\end{array} \right ) }

\newcommand{\su}{{\mathfrak su}}

\newcommand{\la}{\langle}
\newcommand{\ra}{\rangle}

\newcommand{\tr}{{\mathrm Tr}}
\newcommand{\f}{\frac}

\def\nn{\nonumber}
\def\pp{\partial}
\def\arr{\rightarrow}

\def\ka{\kappa}
\def\vphi{\varphi}
\def\eps{\epsilon}

\newcommand{\id}{\mathbb{I}}

\def\bz{\bar{z}}

\def\vV{\vec{V}}
\def\vu{\vec{u}}
\def\vv{\vec{v}}

\def\vsigma{\vec{\sigma}}

\def\tA{\widetilde{A}}
\def\tI{\widetilde{I}}
\def\trho{\widetilde{\rho}}

\begin{document}

\title{Entropy in the Classical and Quantum Polymer Black Hole Models}

\author{{\bf Etera R. Livine}}\email{etera.livine@ens-lyon.fr}
\affiliation{Laboratoire de Physique, ENS Lyon, CNRS-UMR 5672, 46 All\'ee d'Italie, Lyon 69007, France}
\affiliation{Perimeter Institute, 31 Caroline St N, Waterloo ON, Canada N2L 2Y5}

\author{{\bf Daniel R. Terno}}\email{daniel.terno@ mq.edu.au}
\affiliation{Macquarie University, Sydney,  NSW 2109, Australia}

\date{\today}

\begin{abstract}


We investigate the entropy counting for black hole horizons in loop quantum gravity (LQG). We argue that the space of 3d closed polyhedra is the classical counterpart of the space of $\SU(2)$ intertwiners at the quantum level. Then computing the entropy for the boundary horizon amounts to calculating the density of polyhedra or the number of intertwiners at fixed total area. Following the previous work arXiv:1011.5628 \cite{eugenio} we dub these the classical and quantum polymer models for isolated horizons in LQG. We provide exact micro-canonical calculations for both models and we show that the classical counting of polyhedra accounts for most of the features of the intertwiner counting (leading order entropy and log-correction), thus providing us with a simpler model to further investigate correlations and dynamics. To illustrate this, we also produce an exact formula for the dimension of the intertwiner space as a density of ``almost-closed polyhedra".

\end{abstract}

\maketitle

\section{Introduction}


Black holes and their area-entropy law play an essential role in the research of a consistent theory of quantum gravity, both as a basic playground to test each model and as a 
guidance towards understanding the degrees of freedom and the structure of the quantum geometry of space-time.
Indeed, on the one hand, it is necessary that any theory of quantum gravity to recover and explain the entropy, radiation and thermodynamics of black holes, while on the other hand black hole entropy gives us hints that geometry and gravity could be entirely reformulate in terms of quantum information flow.
In the context of loop quantum gravity, black hole horizons are described classically as isolated horizon \cite{isolated}. Then general relativity, expressed in terms of triad and connection variables, induces a Chern-Simons theory on this boundary horizon. First described as a $\U(1)$ Chern-Simons theory in the original work \cite{old}, the analysis was refined and the isolated horizon has been understood to be described by a $\SU(2)$ Chern-Simons theory \cite{alej0} (see also \cite{kaul} for a review of the formalism and of the relation between the $\U(1)$ Chern-Simons theory and the $\SU(2)$ Chern-Simons theory).

At the quantum level, in loop quantum gravity, the state of the geometry of the region outside the black hole is given by a spin network state. This state punctures the horizon and induces topological defects on the surface. These puncture carry $\SU(2)$ representations (spins), which define quanta of area. Then the quantum states of the horizon are then given as states of the Chern-Simons theory, or equivalently as the related conformal blocks. At the mathematical level, this effectively simplifies to describing the horizon states as $\SU(2)$ intertwiner states between the $\SU(2)$ representations carried by the punctures.
Thus, in this context, computing the black hole entropy amounts to counting the number of $\SU(2)$ intertwiners for fixed total area. In this present work, we will discuss neither the conceptual issues of this approach, nor details of the Chern-Simons theory, but we will focus of the entropy counting.

There has already been substantial work on this topic, calculating the number of intertwiners for fixed values of the spins \cite{kaul2,danny} and performing full entropy computations summing over the values of the individual spins for fixed total area \cite{un1,alej0,spanish,alej,mitra} using either exact intertwiner counting or canonical methods. Here, we would like to present an exact intertwiner counting and focus on providing it with an explicit geometrical interpretation.

\smallskip

Based on both the work attempting to understand the deeper structure of the intertwiner space \cite{un0,un1} and research on defining proper coherent intertwiner states \cite{coherent,un2}, a clear picture of intertwiners as quantized polyhedra appeared \cite{un2,polyhedron}. We 
introduce a classical polyhedra counting and we will focus on comparing it to the true quantum intertwiner counting. Our goal is to see how many features of the exact entropy counting can be recovered from a much simpler classical counting. In particular, we will be led to an interesting formula of the number of intertwiners for fixed total area as an integral over almost-polyhedral configurations. This approach allows to see how one deforms from the classical density of states to the number of quantum  states. It will appear that the classical and quantum countings are very similar and for that most purposes the space of classical polyhedra can be a perfect simpler substitute for the space of intertwiners.

More precisely, we describe the space of polyhedra with $N$ faces through the normal vectors to the faces, $\vV_1$,..,$\vV_N$ in $\R^3$. The norm of these vectors give the area of the faces while their direction define the planes in which the faces lay. These vectors are required to satisfy the closure constraint, $\sum_i^N \vV_i=0$. This description of classical polyhedra is particularly suited to building semi-classical intertwiner states in loop quantum gravity \cite{coherent,un2,polyhedron}. It is well-known that such a set of 3-vectors determine a unique (convex) polyhedra, and vice-versa (see \cite{polyhedron} for details on the reconstruction process). As noticed in \cite{eugenio}, this can be seen as a polymer chain, made of $N$ articulated links. This analogy was argued to be especially relevant when discussing the energy or angular momentum of the state. Here we will not investigate those issues, but we will keep the name and thus refer to counting classical polyhedron as the classical polymer model for black holes in loop quantum gravity and to counting intertwiners as its quantum counterpart. We will not use the model introduced in \cite{eugenio} considering classical polyhedra with discrete face area, which can nevertheless be seen as half-way between our purely classical and entirely quantum models.

\smallskip

In the first section, we define precisely our classical polymer model and we describe how to count classical polyhedra for fixed boundary area $A$ and for fixed number of faces $N$. We perform explicitly the sum over $N$ with a chemical potential and we show that the resulting entropy grows at leading order linearly with the area $A$ as expected. We discuss the log-correction to the area-entropy law and show that it is directly related to the closure constraint. We provide explicit closed formula for both the case with and without closure constraint. We show how to compute the fluctuations of the areas of the faces and the correlations between them. We finally show that the number of faces $N$ grows in average linearly with the total area $A$ and that this feature is crucial in recovering the area-entropy law.

The second section reviews and improves the exact intertwiner counting introduced in \cite{un1}. In particular, we show that we recover the classical polyhedra counting as leading order at large area and fixed number of faces. Furthermore, we obtain the same asymptotic formula for the entropy after summing over $N$ up to a shift in the chemical potential.

The third section investigates the relation between the classical and quantum countings and we provide a formula for the dimension of the intertwiner space as counting almost-closed polyhedral configurations, thus reinforcing the interpretation of intertwiners as quantized polyhedra.

Finally, the fourth section is a discussion on the role of the number of faces and the chemical potential. Our main point is that a determination the chemical potential or equivalently the typical ratio between the number of faces and the area will come from an analysis of the coupling of the polyhedra/intertwiners with the exterior geometry. It is the dynamics of our ``horizon" with the geometry outside which will tell us that we are not simply looking at an arbitrary surface but considering an actual black hole horizon.

\section{The Classical Polymer Model}

\subsection{The Micro-Canonical Ensemble and its Phase Space Formulation}
\label{phasespace}


The ensemble that we are interested in is the set of convex polyhedra with $N$ faces and fixed total area $A$. A convenient description for our purpose is through the normal vectors $\vV_i\in\R^3$ to the faces, where $i=1..N$ labels the face: the vector $\vV_i$ is perpendicular to the plane of the face and its norm gives twice the area $|\vV_i|$ of the face. The total area is then $A=\f12\sum_i |\vV_i|$.
The only constraint that these normal vectors have to satisfy is a closure constraint:
\be
\sum_i \vV_i =0\,.
\ee
Such   set of vectors 
determines uniquely a convex polyhedron (embedded in $\R^3$). The interested reader will find more 
details in \cite{polyhedron} with Minkowski's uniqueness theorem and Lasserre reconstruction algorithm. In the present work, we will not describe the reconstruction process of the polyhedron from the normal vector data. We will focus on evaluating the number of such convex polyhedra and use it to define the black hole entropy in the context of loop quantum gravity.

We call this the {\it classical polymer model} for black holes, following the previous work \cite{eugenio}. Indeed, instead of interpreting the $\vV_i$'s as the vectors normals to the polyhedron's faces, one can see them as a linear sequence of vectors forming a loop coming back to its starting point (due to the closure constraint) and of given total length (fixed total area), thus forming some kind of polymer chain formed by $N$ elements.

\medskip

Before moving on, we would like to describe the phase space formulation of this ensemble of closed polyhedra with $N$ faces. This will become essential when quantizing polyhedra to $\SU(2)$ intertwiners and describing the micro-canonical ensemble for black holes in loop quantum gravity.

Following \cite{spinor,un4,merce}, we introduce $N$ spinor variables $z_i\in\C^2$, which are complex vectors living in the fundamental representation of $\SU(2)$:
$$
|z\ra =\mat{c}{z^0 \\ z^1} \quad\in\C^2\,.
$$
Each such spinor determines a unique 3-vector $\vV(z)\in\R^3$ by projecting the 2$\times$2 matrix $|z\ra\la z|$ on the Pauli matrices $\sigma^a$:
\be
\vV
\,\equiv\,
\la z|\vsigma| z\ra,
\quad
|\vV|=\la z|z\ra
\qquad
|z\ra\la z|=\f12\left(
|\vV|\id +\vV\cdot\vsigma
\right)\,.
\ee
Conversely, this 3-vector $\vV$ 
determines the original spinor $z$ up to a global $\U(1)$ phase. Indeed, the spinors $e^{i\theta} \,|z\ra$ all give the same vector, or equivalently said $\vV(z)$ is $\U(1)$-invariant.

The next step is to endow the space of spinors with a canonical Poisson bracket \cite{spinor,un4,spinor2,spinor3,merce}:
\be
\{z^A_i,\bz^B_j\}=\,
-i\delta_{ij}\delta^{AB},
\qquad
\{z^A_i,z^B_j\}=\{\bz^A_i,\bz^B_j\}=0\,.
\ee
It is straightforward to check that the components of each of the 3-vectors $\vV_i$ form (decoupled) $\su(2)$ Lie algebra:
\be
\{V_i^a,V_j^b\}=2\delta_{ij}\epsilon^{abc}V_i^c\,.
\ee
Actually the three components $V^a_i$ of the vector $\vV_i$ are the generators of the $\SU(2)$ transformations on the spinor $z_i$.
This is the reason why the quantization of this spinorial phase space will be described in terms of recoupling of $\SU(2)$ representations and that the space of polyhedra will be quantized into the space of $\SU(2)$ intertwiners.

Then the closure constraint $\sum_i \vV_i=0$ can be translated in terms of spinors \cite{un2}:
\be
\sum_i \vV_i=0
\,\Longleftrightarrow\,
\sum_i |z_i\ra\la z_i| = \f12\sum_i \la z_i|z_i\ra\,\id\,.
\ee
The key remark is that these are first class constraints, generating simultaneous $\SU(2)$ transformations on all the spinors $z_i$.

Finally, the fixed total area condition, $\sum_i |\vV_i|=2A$, can also be translated into spinors as
$\sum_i \la z_i|z_i\ra=A$, and generates global $\U(1)$ transformations on all the spinors, i.e simultaneous multiplication by a phase $e^{i\theta}$. This constraint clearly computes with the closure constraints.

At the end of the day, the spinorial phase space $(\C^2)^{N}//\SU(2)\times\U(1)$  obtained by symplectic reduction from the space of $N$ spinors is exactly the space of (convex) framed polyhedra with $N$ faces and fixed total area up to global 3d rotations (and global phase transformations). By symplectic reduction, we mean that we impose the constraints (closure and fixed total area) and that we quotient by the gauge transformations that they generate. By framed polyhedron, we mean that we add the data of a $\U(1)$-phase to each face, which is interpreted as a choice of 2d frame on for each face.

For example, for $N=4$, we are looking at the phase space of framed tetrahedra. An arbitrary  tetrahedron is determined up to 3d rotations by 6 parameters, minus 1 since we are fixing the total area. Then we need to 4 phases, one for each phase, minus 1 since we are quotienting by global phase transformations. In total, this gives $6-1+4-1=8$ parameters, which reproduces as expected the dimension of the reduced phase space $(\C^2)^{4}//\SU(2)\times\U(1)$.

The last main feature of the phase space formalism using spinor variables is the existence of an action of the unitary group $\U(N)$ on the space of framed polyhedra with $N$ faces, which leads to a $\U(N)$-action on the space of intertwiners at the quantum level \cite{un0,un1,un2}. The action is the straightforward $\U(N)$-action as $N\times N$ matrices on the spinors:
\be
|z_i\ra
\quad\longrightarrow\quad
\sum_k U_{ik}\,|z_k\ra\,.
\ee
This action commutes with the closure constraint, leaves the total area invariant and allows to generate any closed polyhedron configuration from a totally squeezed configuration with only two non-trivial faces (with non-vanishing area). Let us point out that this action requires the spinor variables $z_i$ and does not exist if considering solely the normal 3-vectors $\vV_i$. Although the existence of this $\U(N)$-action is very interesting and seems fundamental when investigating the structure and symmetries of the space of intertwiners as explained in \cite{un0,un1,un2}, we will not need it in the present work.

\medskip

This concludes this short review of the phase space formulation of the space of convex polyhedra. We will not explicitly use the spinorial formalism in the present section, in which we will mainly use the vectors $\vV_i$. The spinor variables will become relevant when looking at coherent states for quantum polyhedra and the quantum polymer model for black holes in section \ref{section3}.

\subsection{Counting Polyhedra at Fixed Total Area: Entropy and Asymptotics}
\label{computerho}

Let us now count the number (or more exactly density) of polyhedra with $N$ faces and fixed total area $\sum_i |\vV_i|=2A$:
\be
\label{rho_def}
\rho_N[A]
\,\equiv\,
\lim_{\eps\arr 0}\,
8\pi
\int\prod_i^N\f{e^{-\eps V_i}\,d^3\vV_i}{4\pi V_i}\,
\delta\left(\sum_k V_k-2A\right)\,
\delta^{(3)}\left(\sum_k \vV_k\right)\,,
\ee
where we both impose the closure of the polyhedra and fix the total area. We use the obvious notation $V_k=|\vV_k|$.
The pre-factor $8\pi$ is a normalization to insure tat $\rho_{N=2}=1$, as we will check later. Let us notice that $\rho_{N=1}=0$ by definition.
The measure $d^3\vV/4\pi V$ for each vector is the measure inherited from the analysis of the spinorial phase space and $\SU(2)$ intertwiners \cite{spinor,un4}.
The parameter $\eps>0$ is a regulator\footnotemark that will be sent to 0 at the end of the calculation. It allows to control the behavior of the integrals when the norm of the vectors is sent to $\infty$. In principle, this does not matter since the fixed total area condition $\sum_k V_k=2A$ automatically implies that the norm of each vector is smaller than 2A. Nevertheless, we will Fourier transform the constraints and commute the integrals to perform the calculation and the regulator will become necessary.
\footnotetext{
By using a little trick on the Fourier transform of the $\delta$-distribution, $2\pi\delta(x)=\int dq\,\exp((iq-\eps)x) $ for all values of $\eps\in\C$, we can avoid having to take the limit $\eps\arr 0$ and have an exact expression for the density of state $\rho_N[A]$ for all $\eps\in\R_+$:
$$
\rho_N[A]
\,=\,
8\pi
\int\prod_i^N\f{e^{-\eps V_i}\,d^3\vV_i}{4\pi V_i}\,
\delta\left(\sum_k V_k-2A\right)\,
\delta^{(3)}\left(\sum_k \vV_k\right)
\,=\,
8\pi\,e^{2\eps A}
\int \f{dq}{2\pi} \int\prod_i^N\f{d^3\vV_i}{4\pi V_i}\,
e^{-\eps V_i}e^{iqV_i}\,
\delta^{(3)}\left(\sum_k \vV_k\right)\,,
$$
where the factor $\exp 2\eps A$ will eventually compensate the factor $\exp (-2\eps A)$ resulting from the integral over the Lagrange parameter $q$ in the final expression \eqref{rhoNA}.
}
As we will see below  this regulator plays the same role as the $+i\eps$ shift of the mass in the Feynman propagator in quantum field theory.

Let us Fourier transform both constraints:
\beq
\rho_N[A]
&=&
8\pi
\int\prod_i^N\f{e^{-\eps V_i}\,d^3\vV_i}{4\pi V_i}\,
\int \f{d^3\vu}{(2\pi)^3}\,
e^{i\vu\cdot\sum_k\vV_k}
\int\f{dq}{2\pi}\,e^{iq(\sum_k V_k-2A)}
\nn\\
&=&8\pi\,
\int \f{d^3\vu}{(2\pi)^3}\,\f{dq}{2\pi}\,
e^{-2iqA}
I(q,\vu)^N
\qquad\textrm{with}\quad
I(q,\vu)\equiv\,
\int \f{d^3\vV}{4\pi V}\,e^{-\eps V}e^{iqV}e^{i\vu\cdot\vV}\,. \label{rho-full}
\eeq
The integral $I(q,\vu)$ converges due to the regulator $\eps>0$. We can compute it exactly as a function of $q$ and the norm $|\vu|$ by performing an explicit integration over the vector $\vV$. We first integrate over the angular part of $\vV$ and then over its norm:
\be
I(q,\vu)
\,=\,\int_0^{+\infty} V^2 \f{dV}{V}\,e^{-\eps V}e^{iqV}\,
\int_{\cS^2}\f{d^2\hat{V}}{4\pi}\,e^{i\vu\cdot\vV}
\,=\,\int_0^{+\infty} V dV\,e^{-\eps V}e^{iqV}\,\f{\sin uV}{uV}
\,=\,\f{1}{u^2-(q+i\eps)^2}\,.
\ee
This is exactly as the Feynman propagator in quantum field theory where $\vu$ plays the role of the momentum and $q$ the role of the mass. Therefore we know how to compute the next integral over $\vu$:
\beq
\rho_N[A]&=&
8\pi\,
\int_{\R}\f{dq}{2\pi}\,e^{-2iqA}\int_0^{+\infty} \f{4\pi u^2 du}{(2\pi)^3}\,
\f{1}{(u^2-(q+i\eps)^2)^N}\nn\\
&=&
8\pi\,
\int_{\R}\f{dq}{2\pi}\,e^{-2iqA}\,
\f{1}{2\pi^2}\f{\sqrt{\pi}}{4}\f{\Gamma(N-\f32)}{\Gamma(N)}(\eps-iq)^{3-2N}
\,=\,
\int_{\R}\f{dq}{2\pi}\,e^{-2iqA}\,
\f{(2N-4)!}{(N-1)!(N-2)!}\f{(\eps-iq)^{3-2N}}{2^{2N-4}}\,.
\eeq
The last step is the integration over $q\in\R$, which amounts to the Fourier transform\footnotemark of the inverse monomial $q^{-n}$:
\be
\label{rhoNA}
\rho_N[A]
\,=\,
e^{-2\eps A}\,\f{A^{2N-4}}{(N-1)!(N-2)!}\,.
\ee
\footnotetext{
We use the explicit Fourier transform formula:
$$
\int_{\R} \f{dq}{2\pi}\,\f{e^{-2iqA}}{(\eps-iq)^n}
\,=\,
\f{2^{2n}A^{2n-1}e^{-2\eps A}}{2(2n-1)!}\,.
$$
}
Finally, one can send  the regulator $\eps$ to zero with the pre-factor becoming trivial, $e^{-2\eps A}=1$.

\smallskip

Let us comment on this result.
First, one check that the case $N=2$ gives us as wanted the normalization $\rho_2[A]=1$.
Second, the dependence on the total area $A$ is expected. Indeed, starting with the initial definition \eqref{rho_def}, one can do a change of variable $\vV\arr \vv=\vV/A$ by normalizing the normal vectors by the total area $A$. The change of variable produces an obvious overall pre-factor $A^{2N-4}$ and leaving us with an integral independent from $A$. This area-independent integral leads to the factor $(N-1)!(N-2)!$ depending exclusively on the number of faces $N$.

\medskip

Interpreting this polyhedra ensemble as a model for black holes, we have computed the density of polyhedra with $N$ faces and fixed total boundary area $A$. However, (stationary) black holes are supposed to be described classically by a single parameter, the horizon area $A$ (or equivalently the mass). Therefore, we need to deal with the number of faces. As it was explained in \cite{un1}, there are  two options. One can decide that $N$ is an extra parameter, either a priori independent from $A$ --- some kind of ``quantum hair" as proposed in \cite{alej}, or fixed a posteriori in terms of the area in the case of a physical black hole --- as in \cite{eugenio} as an example. Alternatively, one attempts to sum over the number of faces $N$, in which case 
it is necessary to postulate the relative statistical weight for $N$.

One simple way to reconcile this two points of view is to sum over the number of faces $N$ with a geometric weight $\alpha^N$ and define the corresponding generating functional:
\be
\rho[\alpha,A]
\,\equiv\,
\rho_2[A]+\alpha\rho_3[A]+\alpha^2\rho_4[A]+\dots
\,=\,
\sum_{N\ge 2}\alpha^{N-2}\rho_N[A]
\,=\,
\sum_{N=0}^{\infty} \f{(A\sqrt{\alpha})^{2N}}{N!(N+1!)}
\,=\,
\,\f{I_1(2A\sqrt{\alpha})}{A\sqrt{\alpha}}\,, \label{gcan}
\ee
where $I_1$ is a modified Bessel function of the first kind.

From a mathematical perspective, the parameter $\alpha$ is a weight associated to each face. More precisely, it can be interpreted as a rescaling of the measure $d^3\vV\arr \alpha d^3\vV$ for each normal vector. Since this measure is actually only defined up to a global factor, $\alpha$ simply accounts for this ambiguity.

From a physical point of view, the factor $\alpha^N\,\equiv e^{-\mu N}$ controls the optimal number of faces and $\mu$ can be interpreted as the corresponding chemical potential, as suggested in \cite{alej}. Or we can interpret $\alpha$ as defining a change of area unit with $A\sqrt{\alpha}$ being the physical area corresponding to the algebraic area $A$. This interpretation is compatible with the fact that the factors $\alpha^N$ can be produced by re-defining the integration measure as $d^3\vV\arr \alpha d^3\vV$ as mentioned above.

We define the entropy as usual as the logarithm of the number of states:
\be
S[\alpha,A]\equiv\log\rho[\alpha,A].
\ee
Then one can use the known asymptotic behavior of the Bessel function, or re-derive it by a stationary point approximation as we will show below in section \ref{find_N}, to get the asymptotical behavior of the entropy:
\be
\label{asympt1}
\rho[\alpha,A]
\,\sim\,
\f{1}{2\sqrt{\pi}}\,e^{2A\sqrt{\alpha}}\,\left(
(A\sqrt{\alpha})^{-\f32}-\f3{16}(A\sqrt{\alpha})^{-\f52}+\dots
\right)\,,
\ee
\be
\label{asympt2}
S[\alpha,A]
\,\sim\,
2A\sqrt{\alpha}-\f32\log A +\dots
\ee
On the one hand, we recover a leading order with the entropy proportional to the area, $S\propto A$. As shown in sections \ref{0closure} and \ref{zclosure} where we relax the closure constraints, this leading order is simply due to the density of states for $N$ random vectors with a fixed sum of their norm (fixed total area).
The parameter $\alpha=e^{-\mu}$, interpreted as the chemical potential or a choice of area unit, enters directly this leading order. It allows to change the ratio $\f SA$ between the entropy and area: the entropy depends at leading order on the way we sum over/count the number of punctures.
This is similar to the mechanism proposed in \cite{un1} (and earlier in \cite{bulk}) to fine-tune the ratio $\f SA$ to $\f14$ without changing the Immirzi parameter of loop quantum gravity (defining the fundamental area unit with respect to the Planck scale), but by playing with the statistical parameters defining the sum over the number of faces.

On the other hand, we notice that log-correction comes with the expected $-\f32$ factor, which seems universal. If we were considering $N$ decoupled vectors $\vV_i$ then we would have no log correction. It is the closure constraint $\sum_i \vV_i=0$ that is responsible for the log-correction to the entropy formula. Actually, we have three closure constraints, one for each direction of $\R^3$. Each of them produces a $-\f12$ correction. We illustrate this interpretation in appendix in the sections \ref{0closure} and \ref{zclosure} by computing explicitly  the entropy for a model without closure constraint and a model with closure constraint in a single direction. For instance, if we remove the closure constraint, the log-correction vanishes, and if we impose the closure constraint in a single direction (e.g. z) this leads to a $-\f12$ log-correction. More generally, it seems that having $n$ closure constraints leads to a log-correction with factor $-\f n2$.

On another level, going to the quantum case and counting the dimension of the Hilbert space instead of integrating over classical vectors will affect the leading order, but in a manner which will be under control as we will see below in the section \ref{quantum}. Nevertheless the log-correction will still remain the same.

\subsection{Spinor Tools: Entropy from Gaussian Integrals}

We have computed above the density of classical polyhedra with $N$ faces for fixed total area defined as an integral over $N$ 3-vectors $\vV_i\in\R^3$ satisfying the closure constraint $\sum_i \vV_i=0$. It is interesting to switch back to the spinor variables $z_i\in\C^4$ used to define the phase space structure of the space of polyhedra. Indeed the integrals above over the 3-vectors now become a simple Gaussian integrals over the spinors. This reformulation will be particularly interesting when considering the quantum model, where the intertwiner counting will be also expressed as Gaussian integrals in the spinor variables over coherent intertwiner states.

Taking care of the measure when changing variables, we get the following expression for the density of states:
\beq
\rho_N[A]
&=&
8\pi
\int\prod_i^N\f{d^4z_i}{\pi^2}\,
\delta\left(\sum_k \la z_k|z_k\ra-2A\right)\,
\delta^{(3)}\left(\sum_k \la z_k|\vsigma|z_k\ra\right)\nn\\
&=&
8\pi\,e^{2\eps A}
\int\f{dq}{2\pi}\int \f{d^3\vu}{(2\pi)^3}
e^{-2iqA}\,\int\prod_i^N\f{d^4z_i}{\pi^2}\,
e^{-\sum_k^N \la z_k|(\eps-iq)-i\vu\cdot\vsigma|z_k\ra}\,.
\eeq
We get the same Gaussian integral over each of the spinor variable $z_i$, which is easy to compute:
\be
\int\f{d^4z}{\pi^2}\,
e^{-\la z|(\eps-iq)-i\vu\cdot\vsigma|z\ra}
\,=\,
\f{1}{\det_{2\times 2}\left[(\eps-iq)\id-i\vu\cdot\vsigma\right]}
\,=\,
\f1{\vu^2-(q+i\eps)^2}\,.
\ee
This allows to reproduce the same expression resulting from the integration over the 3-vectors $\vV_i$:
$$
\rho_N[A]
\,=\,
8\pi\,e^{2\eps A}
\int\f{dq}{2\pi}\int \f{d^3\vu}{(2\pi)^3}
e^{-2iqA}\,
\left(\f1{\vu^2-(q+i\eps)^2}\right)^N\,,
$$
and then perform the integrations over the Lagrange multipliers $q$ and $\vu$ as before.

\subsection{Determining the Size  of Faces and Probing the Shape of the Polyhedron}
\label{size_V}

Beyond the mere counting of polyhedra and the resulting entropy for the (classical) polymer model of black hole horizons, one can further investigate the finer properties of the distribution over the space of polyhedra. For instance, what is the typical size of each face? What is the typical shape of the polyhedron?

Singling out one face out of the $N$ faces, we can look at the probability distribution for the norm\footnotemark $V_i$ of the normal vector.
\footnotetext{Since the measure $d^3\vV_i$ is invariant under 3d rotations, only the norm $V_i$ is relevant while the direction $\hat{V}_i$ follows the uniform probability on the 2-sphere.}
This can be derived from the formula \eqref{rho-full} for the density of states by computing the integral $I(q,\vu)$ for the $(N-1)$ other faces while dropping the integral over $V_i$ for our chosen face. Then one can finish the calculation by integrating over $\vu$ and $q$. However, performing these integrals involve complicated integrals of Bessel functions of the second kind. Therefore, we focus on computing the average size of each face and its standard deviation.

Starting with computing  the mean value of the norm $V_i$ of a single normal vector, we insert a factor $V$ in one of the integrals $I(q,\vu)$. This modified integral is easily computed by differentiating $I(q,\vu)$ with respect to $q$:
$$
I(q,\vu)
=
\int \f{d^3\vV}{4\pi V}\,e^{-\eps V}e^{iqV}e^{i\vu\cdot\vV}
=
\f{1}{u^2-(q+i\eps)^2}
$$
\be
\longrightarrow\quad
\tI(q,\vu)
=
\int \f{d^3\vV}{4\pi V}\,V\,e^{-\eps V}e^{iqV}e^{i\vu\cdot\vV}
=
-i\pp_q I(q,\vu)
=
\f{2(\eps-iq)}{(u^2-(q+i\eps)^2)^2}\,.
\ee
Then we can compute the resulting modified distribution by performing the same integrals as before but with slightly different exponents:
$$
\trho_N[A]
\,\equiv\,
8\pi\,
\int \f{d^3\vu}{(2\pi)^3}\,\f{dq}{2\pi}\,
e^{-2iqA}
I(q,\vu)^{N-1}\tI(q,\vu)
\,=\,
e^{-2\eps A}\f{2A^{2N-3}}{N!(N-2)!}\,.
$$
Then we derive the expected the mean size of a single face:
\be
\la V \ra = \f{\trho_N[A]}{\rho_N[A]}
=\f {2A}N\,,
\ee
which is obvious from the constraint $\sum_i V_i=2A$.
Similarly, we extract the variance of the norm $V$ of one normal vector using the same technique:
\be
\widetilde{\tI}(q,\vu)
=
\int \f{d^3\vV}{4\pi V}\,V^2\,e^{-\eps V}e^{iqV}e^{i\vu\cdot\vV}
=
-\pp_q^2 I(q,\vu)
=
\f{-2}{(u^2-(q+i\eps)^2)^2}+\f{8(\eps-iq)^2}{(u^2-(q+i\eps)^2)^3}\,.
\ee
Replacing $I^N$ by $I^{N-1}\widetilde{\tI}$ in the calculation for $\rho_N[A]$, we obtain at the end of the day:
\be
\la V^2 \ra =\f{6A^2}{N(N+1)}
\qquad
\Longrightarrow
\quad
\sqrt{\la V^2 \ra-\la V\ra^2}
\,=\,
\f{A\sqrt{2}}{N}\sqrt{\f{N-2}{N+1}}
\,\underset{N\gg1}\sim\,
\f1{\sqrt{2}}\la V\ra,
\ee
which mean that the probability distribution of the size of a face remains fuzzy as the number of faces $N$ grows.

We can further look at the correlation $\la V_iV_j\ra$ between the size of two distinct faces $i\ne j$. We follow the same steps as previously by replacing $I^N$ by $I^{N-2}\tI^2$. This leads to:
\be
\forall i\ne j,\quad
\la V_iV_j\ra
\,=\,
A^2\f{2(2N-1)}{(N-1)N(N+1)}\,.
\ee
One can check that, by combining this correlation to the variance $\la V_i^2\ra$, we recover the exact relation $\sum_{i,j}\la V_iV_j\ra=4A^2$, as expected due to the constraint $\sum_i V_i=2A$.
Furthermore, we can check that the size of the faces become more and more decoupled as the number of faces $N$ grows:
\be
\la V_iV_j\ra-\la V\ra^2=
-\f{4A^2}{N^2}\,\f{N-2}{2(N^2-1)}
\,\propto\,
\la V\ra^2\,\f1N
\qquad
\Longrightarrow\quad
\f{\sqrt{|\la V_iV_j\ra-\la V\ra^2|}}{\la V\ra}\underset{N\arr\infty}\arr 0\,.
\ee

By further analyzing the correlations between components of the normal vectors, e.g. $\la V^a_i V^b_j \ra$, one would start probing the shape of the polyhedron, but this goes beyond the scope of the present work and we postpone it for future investigation.

\subsection{Determining the Number of Faces}
\label{find_N}

Let us now investigate the probability distribution of the number of faces $N$ itself. Our density of states is given by the series $\sum_N \alpha^{N}\rho_N[A]$. Let us thus analyze the behavior of $\alpha^{N}\rho_N[A]$ in terms of $N$ and see if it has any clear maxima. To this purpose, we use Stirling formula to analyze its behavior at large $N$:
\be
\rho_N[A]\,\underset{N\gg 1}{\sim}\,\f{N^2}{2\pi}\,e^{2N}A^{2N-4}N^{-2N},
\qquad
\alpha^{N}\rho_N[A]
\,\sim\,
\f{N^2}{2\pi A^4}\,e^{2N(1-\log N)+2N\log \tA},\qquad
\tA\equiv A\sqrt{\alpha}\,.
\ee
The exponent $\vphi(N)\,\equiv\,2N(1-\log N)+2N\log \tA$ has a unique fixed point, which is straightforward to determine:
\be
0=\pp_N\vphi=2\log \tA-2\log N  \Leftrightarrow
N_{max}=\tA =A\sqrt{\alpha}.
\ee
Thus we are in linear regime where the number of punctures $N$ grows linearly with the total area $A$. This is consistent with the recent proposals \cite{un2}, \cite{eugenio} and \cite{alej}, and also with older works on black hole entropy in loop quantum gravity (e.g. \cite{older}).

\begin{figure}[htbp]
    \begin{center}
        \resizebox{0.45\textwidth}{!}{\includegraphics{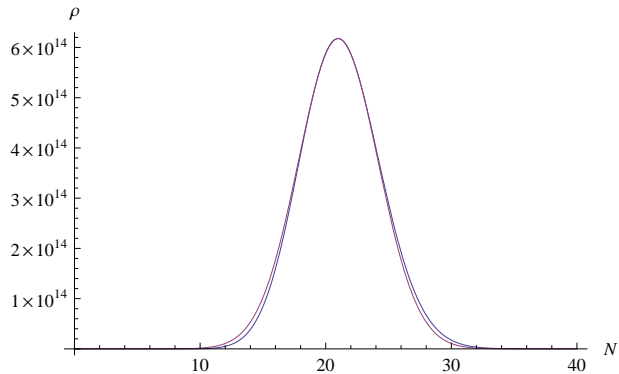}}
    \end{center}
    \caption{\label{choiplot} \small{
Plots of the true density $\alpha^N\rho_N[A]$ for $A\sqrt{\alpha}=20$ and its Gaussian approximation for $N_{max}=A\sqrt{\alpha}+1=21$ (the true density is above the Gaussian approximation). The +1 shift comes from the next-to-leading corrections, which can be easily computed taking into account the $N^2$ pre-factor in the exact density.}}
\end{figure}

Computing the Hessian at the fixed point $\left.\pp^2\vphi\right|_{N=\tA}=-\f2{\tA}$ shows that it is a maxima and that the density $\alpha^{N}\rho_N[A]$ behaves in a first order approximation as a Gaussian around $N_{max}=\tA$. We check numerically the validity of this Gaussian approximation for various values of the parameter. Here figure \ref{choiplot} shows the example of a density peaked around $N_{max}\sim 20$.
Using this Gaussian approximation, we can compute the asymptotics of the the sum over the number of punctures $\sum_N \alpha^{N-2}\rho_N[A]$:
\be
\rho^{approx}[A]
\,=\,
\f1{\alpha^2}\rho_{N_{max}}[A]\,\sqrt{2\pi\f{\tA}{2}}
\,=\,
\f{\tA^2e^{2\tA}}{2\pi \alpha^2A^4}\sqrt{\pi\tA}
\,=\,
\f{1}{2\sqrt{\pi}}\f{e^{2\tA}}{\tA^{\f32}}\,,
\ee
which fits perfectly with the asymptotics \eqref{asympt1}
for $\rho[\alpha,A]$ ad $S[\alpha,A]$.
Let us point out that if we had only considered the evaluation at the optimal number of punctures $\rho_{N_{max}}[A]$, we would have retrieved the correct leading order for the entropy $S[\alpha,A]$ but not the correct log-correction.

\smallskip

Finally, we would like to point out the interesting fact that the entropy $S[\alpha,A]$ is simply given by the optimal number of punctures $N_{max}$ (up to a constant factor 2) at leading order. Although intriguing  at first, this is totally expected from the perspective that the entropy is simply counting the number of degrees of freedom, which is directly proportional to the number of fundamental blocks thus faces at leading order.

\subsection{The Linear Regime $N=\sigma A$}
\label{linear}

Instead of summing over the number of faces $N$, we could take the point of view that it could be fixed by exterior parameters. As we discuss later in section \ref{discuss}, the number of faces $N$ could be fixed dynamically through the coupling of the surface to the external geometry, which would determine if it is a black hole event horizon or an isolated horizon or another type of boundary surface. This scenario was already suggested in \cite{un1}, where the linear regime $N=\sigma A$ with a number of faces increasing linearly with the total boundary area was considered.

Let us start with the density of states $\rho_N[A]$ and define the entropy $s_N[A]\equiv\log \rho_N[A]$. A first tentative regime is defined at fixed number of faces $N$. The asymptotics is obvious to write down:
\be
s_N[A]\,=\,
(2N-4)\log A -\log (N-1)!(N-2)!\,.
\ee
We do not even get a leading order scaling proportionally to the total area $A$. This possibility does not seem to model anything close to a black hole or an isolated horizon.

Following the behavior obtained above for the probability distribution of the number of faces, it seems reasonable to consider instead a linear regime where the number of faces $N$ is determined by the total area $A$ and grows linearly with it. Let us thus set  $N=\sigma A$ with a fixed parameter $\sigma$. It is fairly easy to extract the asymptotics of the resulting entropy using the Stirling formula for the factorials:
\be
\rho_{N=\sigma A}[A]
\,\underset{A\arr\infty}{\sim}\,
\f{\sigma^2}{2\pi A^2}\,\left(\f e\sigma\right)^{2\sigma A}
\qquad\longrightarrow\quad
s_\sigma[A]\equiv\log \rho_{N=\sigma A}[A]
\,\underset{A\arr\infty}{\sim}\,
2\sigma(1-\log\sigma)A - 2\log A+\dots
\ee
We do recover the wanted area-entropy law with the leading order ratio $s/A$ depending explicitly on the ratio $\sigma=N/A$ between the number of faces and the area. A suggestion of \cite{un1} is to use this freedom in the linear regime to fine-tune the ratio $s/A$ to the appropriate physical value by playing with $\sigma$ and without having to fix the Immirzi parameter of loop quantum gravity (choice of physical area unit) to a specific value. In this scenario, we hope that the dynamics induced by the coupling of our surface with the external geometry will fix the ratio $\sigma$ to some appropriate value.

A slightly negative point is that the log-correction does not come with the expected $-\f32$ factor. Nevertheless, this issue is easily solved by allowing Gaussian fluctuations of the number of faces $N$ around the optimal value $\sigma A$. This is exactly what we have seen earlier when summing over the number of faces with the ``trivial" weight $\alpha^N$ defined in terms of the chemical potential. We then recover the expected $-\f32$ log-correction.

Finally, if we look at a regime where $N$ grows much faster than $A$, we lose the linear behavior of the entropy. For instance, considering a power law growth $N\sim A^s$ with $s>1$, the factorials $N!^2$ at the denominator of $\rho_N[A]$ will dominate over the numerator growing as $A^{2N}$ and the density of states will decrease as the area grows.

\medskip

Thus the only viable regime to recover the area-entropy law is the linear regime where the number of faces $N$ grows linearly with the surface area $A$. This supports considering the sum over $N$ with the simple weight $\alpha^N$, which reproduces this exact regime at leading order and indeed predicts a probability distribution for $N$ peaked around an optimal value growing linearly with the area $A$.

\section{The Quantum Polymer Model: Counting Intertwiners}
\label{quantum}

\subsection{Computing the Dimension of the Intertwiner Space and Generating Functionals}
\label{counting_intertwiner}

At the quantum level, the black hole entropy in Loop Quantum Gravity is given by counting the number of $\SU(2)$ intertwiners for a fixed total area \cite{alej0}. Considering intertwiners with $N$ legs, we put one spin $j_i$ on each leg labeled by $i=1..N$ and we look at the $\SU(2)$-invariant states in the tensor product of the corresponding irreducible representations:
\be
\cH^N_{j_1,..,j_N}
\,\equiv\,
\textrm{Inv}_{\SU(2)} \cV^{j_1}\otimes..\otimes \cV^{j_N}\,,
\ee
where $\cV^j$ stands for the $(2j+1)$-dimensional Hilbert space carrying the irreducible representation of $\SU(2)$ of spin $j$.

As was shown in \cite{un1,un2,spinor}, and then reviewed in \cite{un4,polyhedron}, this space of intertwiners comes as the quantization of the space of classical polyhedra as defined in section \ref{phasespace}; more precisely it is the space of holomorphic $L^2$ functions on the spinorial phase space. From this point of view, we interpret intertwiners as quantized polyhedra and we see the LQG black hole entropy counting of intertwiners as the quantum counterpart of the polymer model defined and analyzed in the previous section. For more details on the equivalence between the intertwiner space and quantized polyhedra, the interested reader can refer to \cite{un1,un2,spinor,un4} and to \cite{polyhedron}.

We would also like to cite \cite{eugenio}, which proposes an intermediate model between counting classical polyhedra and quantum intertwiners. It quantizes the area of the faces while keeping the direction of their normal vector freely distributed on the classical 2-sphere. Although this model is not directly relevant to our present work, we give its basic definition and a rough analysis in appendix \ref{BS}.

\medskip

Next, we define the Hilbert space of $\SU(2)$ intertwiners for a fixed total area, summing over spin labels $j_i$ while keeping their sum fixed:
\be
\cH^N_J
\,=\,
\bigoplus_{\sum_i^Nj_i=J}
\cH^N_{j_1,..,j_N}
\,=\,
\bigoplus_{\sum_i^Nj_i=J}
\textrm{Inv}_{\SU(2)} \cV^{j_1}\otimes..\otimes \cV^{j_N}\,.
\ee
Let us point out that we are using here the equidistant area spectrum of Loop Quantum Gravity, where the area of a face carrying a spin $j$ is given actually by $j$ in Planck unit (times the Immirzi parameter). This is in contrast with the more used spectrum given by the square-root of the $\SU(2)$ Casimir operator, $\sqrt{j(j+1)}$ in Planck unit. Let us remark that these two spectra only differ by operator-ordering ambiguities (see e.g. \cite{area_krasnov}). In our context, the equidistant spectrum $\cA_j\propto j$ comes naturally from the quantization of the spinorial phase space leading to the space of intertwiners\footnotemark. In particular, it allows to realize that the intertwiner space $\cH^N_J$ at fixed total area $J=\sum_i j_i$ carries an irreducible representation of the unitary group $\U(N)$, which is then interpreted as the group of gauge-invariant deformations of $N$-legged intertwiners at fixed area \cite{un1,un2}.
\footnotetext{
This equidistant spectrum  also turns out to be useful and relevant in other contexts. For instance, in the construction of the EPRL-FK spinfoam models with Immirzi parameter, one needs to use the equidistant spectrum to solve exactly the diagonal simplicity constraints \cite{eprl-fk}.
}

The dimension of the space of intertwiners $\cH^N_J$ was computed in \cite{un1} actually using the fact that it carries an irreducible representation of $\U(N)$ with a two-line Young tableau:
\be
\label{undim}
d_N[J]
\,\equiv\,
\dim \cH^N_J
\,=\,
\f1{J+1}\bin{J}{N+J-1}\bin{J}{N+J-2}\,,
\ee
in terms of binomial coefficients. As also shown in \cite{un1}, this dimension can also be computed as the integral over $\SU(2)$ of the product of suitable characters:
\be
\label{int_character}
d_N[J]
\,=\,
\int_{\SU(2)} dg\,
\sum_{\sum_i^N j_i=J} \prod_i^N \chi_{j_i}(g),
\ee
where $dg$ is the normalized Haar measure and $\chi_j$ the $\SU(2)$-character of spin $j$ (i.e the trace of the matrix representing the group element $g$ in the representation of spin $j$).

For a fixed number of faces $N$  and a large area $J\arr\infty$ we get 
\be
d_N[J]
\,\underset{J\arr\infty}{\sim}\,
\f{J^{2N-4}}{(N-1)!(N-2)!}+\frac{NJ^{2N-5}}{(N-1)!(N-3)!}+\ldots \label{expan-2}
\ee
which in the leading order is the same as the classical density of states $\rho_N[J]$ derived in \eqref{rhoNA}. Thus the number of states $d_N[J]$ and the density of states $\rho_N[A]$ at fixed number of faces $N$ coincide for large areas is consistent with seeing the intertwiner counting as the quantized version of the classical polymer black hole model counting classical polyhedra.
This is a strong result showing the consistency of the approach of counting classical polyhedra for modeling the isolated horizon state counting in loop quantum gravity at least at leading order.
As a consequence, we expect very similar results for the generating functional and the resulting entropy.

The next step is to sum over the number of legs $N$ (or equivalently the number). At fixed total area $J$, we sum over $N$ with a geometric weight $\alpha$ and obtain a hypergeometric function:
\be
d[\alpha,J]=\sum_{N\ge 2} \alpha^{N-2}d_N[J]
\,=\,
\sum_{N=0}^\infty \alpha^N\f{(N+J)!(N+J+1)!}{N!(N+1)!J!(J+1)!}
\,=\,
{}_2F_1(J+1,J+2,2,\alpha).
\ee
As shown in \cite{un1}, or using the known asymptotics for the hypergeometric functions, or working out a saddle point approximation for the above series, one can extract the large area behavior of this generating functional:
\be
\label{entropy_asympt}
\cS[\alpha,J]\equiv\log d[\alpha,J]
\underset{J\gg 1}{\sim}\,
-2J\log(1-\sqrt{\alpha})-\f32\log J\,.
\ee
Thus we find that classical polyhedra and quantum intertwiners lead to very similar entropies, which differ only in a renormalization of the weight $\alpha$ associated to each puncture (or equivalently the chemical potential):
\be
S[\alpha,A]\sim 2A\sqrt{\alpha}-\f32\log A
\quad\longrightarrow\quad
\cS[\alpha,J]\sim
-2J\log(1-\sqrt{\alpha})-\f32\log J.
\ee
Identifying the total area $A=J$ with the sum of the spins, we see that the ratio $S/A$ differs between the classical model and the exact quantum computation. This change of leading order between the classical polymer and quantum polymer is due to the quantization and subsequent discretization of the area spectrum. Both entropies have the same leading order when the weight associated to each puncture is very small, $\alpha\arr 0$.
Moreover, simply renormalizing the chemical potential by changing $\sqrt{\alpha}$ into $-\log(1-\sqrt{\alpha})$, allows to get the quantum entropy from the classical entropy formula. We will 
this classical $\leftrightarrow$ quantum transition later in Sec.~\ref{section3}.
Let us also point out that the quantum entropy $\cS[\alpha,J]$ is always smaller than its classical counterpart $S[\alpha,J]$.

Moreover an interesting point is that the log-correction is exactly the same in both cases. This robustness of the log-correction can be interpreted as  coming from the number of constraints that we are imposing: the closure constraints in $\R^3$ in the classical case corresponds to the requirement of $\SU(2)$-invariance in the quantum theory. Since $\SU(2)$ is also 3-dimensional, the $\SU(2)$-invariance translates into 3-dimensional constraints, which lead to the factor $-\f32$. We illustrate this by working out explicitly in appendix \ref{relax} the entropy counting for models without $\SU(2)$-invariance (where there will be no log-correction) and with a simpler $\U(1)$-invariance (where we recover a $-\f12$ log-correction).

Finally, using the Stirling approximation for the factorials for large $J$ and $N$, we can compute the optimal number of faces $N_{max}$ given fixed total area $J$ and chemical potential $\alpha$. Identifying the maximum of the probability distribution $\alpha^{N}d_N[J]$ easily gives at leading order:
\be
N_{max}\sim \f{J\sqrt{\alpha}}{1-\sqrt{\alpha}},
\ee
which is almost the same as the classical maximum up to the factor $1-\sqrt{\alpha}$ in the denominator (which nevertheless becomes trivial if $\alpha$ is sent to 0).

\subsection{Computing the Generating Functional}

Let us look  into defining the generating functionals for the dimensions $d_N[J]$. This will be particularly useful when comparing the classical and quantum entropies using coherent states in the next section.
Following \cite{un1}, we can define a generating functional $f_N[t]=\sum_{J\in\N} t^{2J} d_N[J]$ by summing over the total area $J$. We can express it as an integral over $\SU(2)$ using the formula \eqref{int_character} for $d_N[J]$:
\be
\label{fN_def}
d_N[J]
\,=\,
\int_{\SU(2)} dg\,
\sum_{\sum_i^N j_i=J} \prod_i^N \chi_{j_i}(g)
\quad\arr\quad
f_N[t]
\,\equiv\,
\sum_{J\in\N} t^{2J} d_N[J]
\,=\,
\int_{\SU(2)}dg\,\left(
\sum_{j\in\N/2} t^{2j}\chi^{j}(g)
\right)^N
\ee
Taking into account that the characters are class functions (invariant under conjugation) and parameterizing the group element $g$ in terms of its class angle $\phi$, we write the generating functionals $f_N[t]$ as integrals over the single angle $\phi$:
\be
\label{fN_int}
f_N[t]
\,=\,
\f2\pi\int_0^{\pi}\sin^2\phi d\phi\,\left(
\sum_{j\in\N/2} t^{2j}\f{\sin(2j+1)\phi}{\sin\phi}
\right)^N
\,=\,
\f2\pi\int_0^{\pi} d\phi\,
\f{\sin^2\phi}{(t^2-2t\cos\phi+1)^N}\,.
\ee
Then we can introduce an overall generating functional by also summing over the number of faces $N$:
\be
F[\alpha,t]=\sum_{N\in\N} \alpha^{N}f_N[t]
=\sum_J t^{2J} \sum_{N\in\N} \alpha^{N}d_N[J]
=\sum_J t^{2J} (1+\alpha^2 d[\alpha,J]),
\ee
which is thus also a generating functional for the dimensions $d[\alpha,J]$.
Explicitly computing the sum over $N$ allows to write $F[\alpha,t]$ as a trigonometric integral:
\be
\label{F_int}
F[\alpha,t]=
\f2\pi\int_0^{\pi} \sin^2\phi d\phi\,
\f{(t^2-2t\cos\phi+1)}{(t^2-2t\cos\phi+1)-\alpha}\,.
\ee
This integral can be computed explicitly from $\alpha,t\sim 0$ \cite{un1}:
\be
F(\alpha,t)
\,=\,
1+\f\alpha2+\f\alpha{2t^2}
\left(
1-\alpha-\sqrt{(\alpha-(1+t)^2)(\alpha-(1-t)^2)}
\right)
\ee
For $0< \alpha <1$, this generating functional is well-defined for $t$ around 0. Then the lowest (non-vanishing) pole is at $t_c=1-\sqrt{\alpha}$, or equivalently at $\alpha=(1-t_c)^2$. This gives us the asymptotic behavior for the coefficients of the power series in $t^{2J}$ \cite{un1}:
\be
\log \sum_{N\in\N} \alpha^{N}d_N[J]
\,\underset{J\arr\infty}{\sim}\,
-2J\log t_c = -2J\log(1-\sqrt{\alpha}),
\ee
which reproduces exactly the leading order of the asymptotics \eqref{entropy_asympt} of the entropy.

\subsection{A Remark on the ``No-Trivial Puncture" Model and Entropy}

An issue with the intertwiner counting presented above in section \eqref{counting_intertwiner} is that it includes vanishing spins on the intertwiner legs $j_i=0$ \cite{un1}. This corresponds to faces with vanishing area. Therefore, one would like to ideally remove them and perform a ``no trivial-spin" counting. As was shown in \cite{un1} and will be explained below, the ``no trivial-spin" entropy counting can be re-absorbed in a re-definition of the weight (chemical potential) $\alpha$. Indeed removing trivial spins amounts simply to a specific summing over the number of faces $N$.
Thus all one needs is the generating functional $d[\alpha,J]$ summing over the number of punctures $N$ and this density of states is enough to take into account removing 0-spins or even more general restrictions one would like to impose on the intertwiner counting. In particular, the ``no trivial-spin" entropy $\cS_{\varnothing}[\alpha,J]$ will have the same asymptotical behavior as $\cS[\alpha,J]$, with a renormalized factor for the leading order and the same log-correction.

Before showing the details of the computation of this ``no trivial-spin" entropy $\cS_{\varnothing}[\alpha,J]$, we would like to discuss the legitimacy of removing the contributions of 0-spins. From a classical and geometrical point of view, it is natural to remove faces with vanishing area, since these are simply not actual faces. From a quantum point of view, removing 0-spins amounts to removing the ground state of the Hilbert space and the legitimacy of doing so can be questioned. First, considered solely the horizon, there is no principle or reason to exclude quantum superposition of spins including 0-spin on a face/leg of the intertwiner. Second, considering the coupling of the black hole horizon with the exterior space-time geometry, the number of faces $N$ is a property of the quantum state of geometry of the external region and should come from the edges of this spin network state puncturing the black hole horizon. Nothing forbids these edges to carry superpositions of various spins including 0-spin. From this point of view, the number of faces $N$ should be dynamically derived from the coupling of the black hole with the external geometry. However, to go further in this direction would require a kinematical and dynamical model of the near-horizon geometry in loop quantum geometry and it is not enough to have a kinematical picture of the horizon decoupled from the rest of the space-time geometry.

\medskip

The number of intertwiner $D_N[J]$ for fixed total area $J$ with a number of faces $N$ constrained to carry non-vanishing spin is easily related to the number of intertwiners $d_N[J]$ with possible 0-spins through the binomial transform, as used in \cite{un1}:
\be
d_N[J]=\sum_{K=0}^N\mat{c}{N\\K}\,D_K[J],
\qquad
D_K[J]=\sum_{N=0}^K(-1)^{K-N}\,\mat{c}{K\\N}\,d_N[J],
\ee
in terms of binomial coefficients. Then using the series expansion\footnotemark of $(1-x)^{-N}$, one can easily relate the two generating functionals:
\footnotetext{
We use the following series expansion around $x=0$:
$$
\f{x^N}{(1-x)^{N+1}}=\sum_{K\ge N}x^K\,\mat{c}{K\\N}\,.
$$
}
$$
\sum_N\alpha^N d_N[J]
\quad\overset{?}{\longrightarrow}\quad
\sum_N\alpha^N D_N[J]
$$
\be
\sum_{K=0}^\infty\alpha^K D_K[J]
\,=\,
\sum_{N=0}^\infty d_N[J]\sum_{K=N}^\infty (-1)^{K-N}\alpha^K\,\mat{c}{K\\N}
\,=\,
\f1{1+\alpha}\sum_N \left(\f\alpha{1+\alpha}\right)^N d_N[J]\,.
\ee
Therefore, summing over faces carrying non-trivial spins is fully equivalent to summing over faces with arbitrary spins, up to a controlled renormalization of the weight $\alpha$ of individual faces.

Thus, introducing the non-trivial spin entropy $\cS_{\varnothing}[\alpha,J]\,\equiv\,\log\sum_{N\ge2}\alpha^{N-2} D_N[J]$, where we removed the nonsensical case $N=0$ and the vanishing term $D_1[J]=0$, its asymptotics at large area $J\gg 1$ is automatically derived from the known behavior  of the entropy $\cS[\alpha,J]$ given by \eqref{entropy_asympt}:
\be
\cS_{\varnothing}[\alpha,J]
\underset{J\gg 1}{\sim}\,
-2J\log\left(1-\sqrt{\f\alpha{1+\alpha}}\right)-\f32\log J+\dots\,.
\ee
To conclude with the issue of removing trivial faces or not, we do not need to consider explicitly the case of removing the contribution of 0-spins since this is a possibility already taken in account by considering the generic generating functional $d[\alpha,J]$ summing over the number of faces $N$.

\section{Bridging between the Classical and Quantum Models}
\label{section3}

The Hilbert space of $\SU(2)$ intertwiners with $N$ legs is understood to be the quantization of the space of closed polyhedra. It would thus be enlightening to understand how the intertwiner counting can be actually interpreted as computing a number of classical polyhedra or ``almost-classical" polyhedral configurations. To this purpose, we will use coherent intertwiner states to write a semi-classical formula for the dimensions of the intertwiner spaces as an integral over fuzzy polyhedra.
This would allow to bridge between the entropy formula for the classical and quantum polymer models described previously.
Before getting into the details of coherent states, let us start with a simple remark on the entropy counting. Indeed, as we already pointed out earlier, it appears that we can obtain the classical entropy formula as a double scaling limit of the quantum entropy formula sending both the area to infinity and the chemical potential to 0. For fixed parameters $A$ and $\tilde{\alpha}$, we set $J=\ka A$ and $\alpha=\ka^{-2}\tilde{\alpha}$ in the intertwiner counting and send the scaling parameter $\ka$ to $\infty$:
\be
\cS[\alpha,J]\underset{J\gg 1}{\sim} -2J\log(1-\sqrt{\alpha})\quad
\longrightarrow
\cS[\ka^{-2}\tilde{\alpha},\ka A]\underset{\ka\arr\infty}{\sim}
-2\ka A\log(1-\ka^{-1}\sqrt{\tilde{\alpha}})\sim 2A\sqrt{\tilde{\alpha}}
\sim S[\tilde{\alpha},A]
\ee
Let us point out this double scaling limit is the one that keeps fixed the most probable number of faces $N_{max}$ of the intertwiner/polyhedron. Thus in a sense, we rescale the chemical potential at the same time as the area in order to keep the number of faces finite.

This suggests that we should be able to recover the classical polymer model in a certain large area regime of the quantum model, or vice-versa 
we should be able to interpret the quantum polymer model as a suitable deformation of the classical model.

\subsection{Semi-Classical Formula for Intertwiners and Quantum Polyhedra}

Here we would like to propose an alternative expression for the number of intertwiners as an integral over classical vectors representing fuzzy or actual polyhedra.
Following the approach of \cite{coherent,polyhedron} to define coherent intertwiners through group averaging, we can introduce the $\SU(2)$ coherent states $|j,z\ra$ labeled by spinors $z\in\C^2$ living in the fundamental representation of $\SU(2)$. Those $|j,z\ra$ states form a over-complete basis of the irreducible representation $\cV^j$ of spin $j$ and transform covariantly under $\SU(2)$ (see e.g. \cite{un2,un4,spinor,spinor2} for details on their definition and properties).
Here, we will use the decomposition on the identity on the representation $\cV^j$ of spin $j$:
\be
\id_{\cV_j}
= \f1{(2j)!}\int \f{e^{-\la z | z\ra }d^4z}{\pi^2}\,
|j,z\ra\la j,z|\,,
\ee
where  $\la z | z\ra $ stands for the norm of the spinor $z$ in $\C^2$.
This allows to express the characters $\chi_j(g)$ as Gaussian integrals over $\C^2$:
\be
\chi_j(g)
\,=\,
\f1{(2j)!}\int \f{e^{-\la z | z\ra }d^4z}{\pi^2}\,
\la j,z|g|j,z\ra
\,=\,
\f1{(2j)!}\int \f{e^{-\la z | z\ra }d^4z}{\pi^2}\,
\la z|g|z\ra^{2j}\,,
\ee
where $g|z\ra$ is the natural action of the group element $g$ on the complex spinor $z$ in the fundamental representation of $\SU(2)$. We can insert this expression in the definition of the dimension $d_N[J]$ as an integral over $\SU(2)$ of a product of characters:
\be
\label{coh1}
d_N[J]
\,=\,
\int_{\SU(2)} dg\,
\sum_{\sum_i^N j_i=J} \prod_i^N \chi_{j_i}(g)
\,=\,
\int \prod_i^N\f{e^{-\la z_i | z_i\ra }d^4z_i}{\pi^2}
\sum_{\sum_i^N j_i=J}\f1{\prod_i(2j_i)!}
\int dg\, \la z_i|g|z_i\ra^{2j_i}\,.
\ee
We can write this integral in term of the 3-vectors $\vV_i=\la z_i|\vsigma|z_i\ra$ (since the expression is invariant under multiplication by a phase of each spinor $z_i$). Parameterizing explicitly the group element as $g=\cos\theta+i\sin\theta \hat{u}\cdot\vsigma$ in term of the class angle $\theta\in[0,\pi]$ and the rotation axis $\hat{u}\in\cS_2$, we have:
\be
d_N[J]
\,=\,
\int \prod_i^N\f{e^{-V_i}d^3\vV_i}{4\pi V_i}
\sum_{\sum_i^N j_i=J}\f1{\prod_i(2j_i)!}
\f2\pi\int \sin^2\theta d\theta \int_{\cS_2}\f{d^2\hat{u}}{4\pi}\,
(V_i\cos\theta+i\sin\theta \hat{u}\cdot\vV_i)^{2j_i}\,.
\ee
Besides the sum over the spins $j_i$, this is an entirely classical expression for the dimension of the intertwiner space. Following the original approach of \cite{coherent}, one can do a saddle point analysis of the integral over $\SU(2)$ for large spins $j_i$, in which case one finds that this integral is exponentially suppressed except if the closure constraint $\sum_i j_i\vV_i=0$ is satisfied. This is not the closure constraint of the classical model. We can nevertheless restrict the integration of over the coherent states to unit spinors, $\la z_i|z_i\ra=1$, and thus to unit vectors $\hat{v}_i\in\cS_2$ which we write with a small $v$ to distinguish them from the normal vectors $\vV_i$ with arbitrary norm. These still provide a decomposition of the identity on the $\SU(2)$ representation spaces $\cV^j$. This are actually the procedure used in \cite{coherent,polyhedron} when defining coherent intertwiner states. Then the closure constraint determining the saddle point at large spins is $\sum_i j_i\hat{v}_i=0$, which can be written simply as $\sum_i \vV_i=0$ after defining the normal vectors as $\vV_i\equiv j_i \hat{v}_i$ with $V_i\equiv j_i$ and thus $\sum_i V_i =\sum_i j_i=J$. The problem with this formula is two-fold.
First, the norm of the normal vectors are quantized and discrete and we do not have an integral over classical continuous vectors in $\R^3$. This corresponds more to the polymer model by Bianchi \`a la Bohr-Sommerfeld as presented above in section \ref{BS}.
Second, the integrand does not admit a simple natuural expression in terms of these normal vectors $\vV_i$ with discrete norm equal to $j_i$.
Instead, we would like to propose here another semi-classical formula for the intertwiner counting based on improved coherent intertwiner states, as defined in \cite{un2,un4}, which are constructed as $\U(N)$ coherent states.

\smallskip

Indeed, it was proven in \cite{un1} that the Hilbert space of intertwiners with $N$ legs and fixed total spin $J=\sum_i j_i$ is an irreducible representation of the unitary group $\U(N)$:
$$
\cH_J^N=\bigoplus_{\sum_i j_i=J}
\textrm{Inv}_{\SU(2)} V^{j_1}\otimes..\otimes V^{j^N}.
$$
It was then proposed in \cite{un2} to define coherent intertwiner states in $\cH_N^J$ consistent with the $\U(N)$ action. Here we do not need the $\U(N)$ action, we will thus refer the interested reader to \cite{un2,un4} for details on the definition and analysis of those $\U(N)$ coherent states. We will simply use the formula for the decomposition of the identity on the Hilbert space $\cH_N^J$:
\be
\id_N^J
\,=\,
\f{1}{{J!(J+1)!}}
\int \prod_i\f{d^4 z_i}{\pi^2}
e^{-\sum_i \la z_i|z_i\ra }\,
|J,\{z_i\}\ra\la J,\{z_i\}|\,,
\ee
where the coherent intertwiners are labeled by the total spin $J$ and $N$ spinor variables $z_i\in\C^2$. Those states have a non-trivial norm:
\be
\la J,\{z_i\}|J,\{z_i\}\ra
\,=\,
\left(\det_{2\times 2} \sum_i |z_i\ra\la z_i|\right)^J
\,=\,
\f1{2^{2J}}\,\left(
\sum_{i,j}\la z_j|z_i][z_i|z_j\ra
\right)^J
\,=\,
\f1{2^{2J}}\,\left(
(\sum_i V_i)^2-|\sum_i \vV_i|^2
\right)^J\,.
\ee
This provides a beautifully simple formula for the dimension of the intertwiner space:
\be
d_N[J]=\dim \cH_J^N=\tr \id_N^J
\,=\,
\f{1}{2^{2J}{J!(J+1)!}}
\int \prod_i\f{d^3 V_i}{4\pi V_i}
e^{-\sum_i V_i }
\left((\sum_i V_i)^2-|\sum_i \vV_i|^2
\right)^J\,,
\ee
where we switch the integral over the spinors to an integration over the normal 3-vectors.

\smallskip

It is interesting to check this formula explicitly. We use similar techniques as when computing the classical density of states $\rho_N[A]$. We introduce the Fourier transform parameter $\vu$ and $q$ and write:
\beq
d_N[J]
&=&
\f{1}{2^{2J}{J!(J+1)!}}
\int \prod_i\f{e^{-V_i}d^3 V_i}{4\pi V_i}\,
\left.
(\pp_{u_i}\pp_{u_i}-\pp^2_q)^J\,e^{i\vu\cdot\sum_i \vV_i}e^{iq\sum_i V_i}
\right|_{\vu=q=0}
\nn\\
&=&
\f{1}{2^{2J}{J!(J+1)!}}\,
\left.
(\pp_{u_i}\pp_{u_i}-\pp^2_q)^J\,I(q,\vu)^N
\right|_{\vu=q=0}
\,,
\eeq
where $I(q,\vu)$ is the same integral as computed in section \ref{computerho}:
\be
I(q,\vu)=\int \f{d^3\vV}{4\pi V}\,e^{-\beta V}e^{iqV}e^{i\vu\cdot \vV}
\,=\,
\f1{\vu^2-(q+i\beta)^2},
\ee
where we will take $\beta=1$.
It is easy to compute the action of the Laplacian on the integral kernel:
\be
(\pp_{u_i}\pp_{u_i}-\pp^2_q)\,\f1{\left(\vu^2-(q+i\beta)^2\right)^N}
\,=\,
\f{-4N(N-1)}{\left(\vu^2-(q+i\beta)^2\right)^{N+1}}\,.
\ee
By a straightforward recursion, we get:
\be
(\pp_{u_i}\pp_{u_i}-\pp^2_q)^J\,\f1{\left(\vu^2-(q+i\beta)^2\right)^N}
\,=\,
2^{2J}\,\f{(N+J-1)!(N+J-2)!}{(N-1)!(N-2)!}\,
\f1{\left(\vu^2-(q+i\beta)^2\right)^{N+J}} \,.
\ee
Putting all the pieces together and setting back $\beta=1$, we get:
\be
d_N[J]
\,=\,
\f{(N+J-1)!(N+J-2)!}{J!(J+1)!(N-1)!(N-2)!},
\ee
which is exactly the expected result as given in eqn.\eqref{undim}.
We could have also compute this integral by directly performing the Gaussian integral over the spinor variables, in which case it would have turned into a combinatorial problem of counting the number of pairings in the polynomial $\la J,\{z_i\}|J,\{z_i\}\ra$ using the Wick theorem.

\smallskip

We therefore have a exact formula for the intertwiner counting as an integral over classical 3-vectors $\vV_i$. But we do have neither closure constraints nor a total area constraint. The natural question is whether this integral can be interpreted as an integral over semi-classical configurations or fuzzy polyhedra.

Looking more closely at the factor $((\sum_i V_i)^2-|\sum_i \vV_i|^2)^J$, we can factor out the total area $\sum_i V_i$ and we get a positive expression smaller or equal to 1 to the power $J$:
$$
\left(1-\f{|\sum_i \vV_i|^2}{(\sum_i V_i)^2}\right)^J\,.
$$
Since the normal of the ``closure vector" $\sum_i \vV_i$ is always smaller or equal to the total area $\sum_i V_i$, this expression goes to 0 as $J$ grows except if and only if the closure constraint is satisfied $\sum_i \vV_i=0$. Thus our integral is clearly peaked on closed configuration and more so as the total spin $J$ grows. And we do have an integral over some kind of fuzzy polyhedra. We can be slightly more precise and write:
\be
((\sum_i V_i)^2-|\sum_i \vV_i|^2)^J
\,=\,
(\sum_i V_i)^{2J}\,e^{J\ln \left(1-\f{|\sum_i \vV_i|^2}{(\sum_i V_i)^2}\right)}
\,\sim\,
(\sum_i V_i)^{2J}\,e^{-J\,\f{|\sum_i \vV_i|^2}{(\sum_i V_i)^2}}\,,
\ee
where it becomes clear that we have approximatively at large area $J$ a Gaussian peaked on the closure  constraint for classical  polyhedra.

\smallskip

On the other hand, the total area $\sum_i V_i$ does not seem to be fixed. In order to find out, let us compute the average of  $\sum_i V_i$ and its fluctuations. We can compute directly  the average of $e^{\eta \sum_i V_i}$ and this will contain the average and all the higher moments. This does not require any further calculation since it amounts to taking $\beta =(1-\eta)$ and plugging it in the expressions above. This leads to:
\beq
\la e^{\eta \sum_i V_i} \ra
&\equiv&
\f{1}{d_N[J]2^{2J}{J!(J+1)!}}
\int \prod_i\f{d^3 V_i}{4\pi}
e^{-(1-\eta)\sum_i V_i }
\left((\sum_i V_i)^2-|\sum_i \vV_i|^2
\right)^J \nn\\
&=&
\f1{(1-\eta)^{N+J}}
\,=\,
\sum_{n\in\N}
\f{\eta^n}{n!}\,\f{(n+N+J-1)!}{(N+J-1)!}\,.
\eeq
This perturbative expansion allows to read the average and the fluctuation:
\be
\la \sum_i V_i \ra=(J+N),
\qquad
\f{\sqrt{\la (\sum_i V_i)^2 \ra -\la \sum_i V_i \ra^2}}{\la \sum_i V_i \ra}
\,=\,
\f1{\sqrt{J+N}}\,.
\ee
Thus the total area $\sum_i V_i $ is peaked as the total spin $J$ grows, but its average value has an intriguing $+N$ shift.

At the end of the day, we do have a semi-classical formula for the intertwiner counting as an integral over the normal 3-vectors representing some fuzzy polyhedra almost satisfying the closure constraint and almost of fixed total area. This provides a simple geometric interpretation to the counting of intertwiners.
Using this semi-classical formula, one could now go further and  compute all the averages and correlations for the normal 3-vectors in the quantum polymer model.

\subsection{Gauge-Fixing to Closed Polyhedra}

Up to now, we have written the dimension of the intertwiner space as an integral over almost-closed configurations representing some kind of fuzzy polyhedra. It is actually possible to gauge-fix this integral to exactly closed polyhedral configurations.
Indeed following the previous work on coherent intertwiners \cite{un2,un4}, it turns out that the coherent intertwiner states are invariant under global $\SL(2,\C)$ transformations on the spinor labels $z_i$, while this affects the closure constraint:
\be
|z_i\ra\in\C^2\arr\Lambda\,|z_i\ra,
\qquad
|J,\{z_i\}\ra =|J,\{\Lambda\,z_i\}\ra,\qquad
|z_i\ra\la z_i|\arr \Lambda\,|z_i\ra\la z_i|\,\Lambda^\dag\,,
\ee
where $\Lambda\in\SL(2,\C)$ acts as a 2$\times$2 matrix.
The key point is that, for an arbitrary set\footnotemark of spinors $z_i$, it is always possible to find a transformation $\Lambda$ such that the transformed spinors satisfy the closure constraint, $\Lambda\,\sum_i|z_i\ra\la z_i|\,\Lambda^\dag\propto\id$. Indeed, as noticed in \cite{un2}, the  2$\times$2 matrix $X\equiv \sum_i|z_i\ra\la z_i|$ is Hermitian and strictly positive and can always be written as $X=\lambda \Lambda^{-1}\Lambda^{-1}{}^\dagger$ with $\lambda>0$ and $\Lambda\in\SL(2,\C)$.
\footnotetext{
The only exception is the degenerate case when all the spinors are proportional to each other. In this case, the  2$\times$2 matrix $X=\sum_i|z_i\ra\la z_i|$ has a vanishing determinant and there does not exist a $\SL(2,\C)$ matrix mapping this degenerate set of spinors on a closed configuration. However, this set is of measure zero and does not play any important role in our integrals.
}
Reversely, starting with arbitrary closed configuration, we can obtain any random (non-closed) configuration by $\SL(2,\C)$ transformations. This way, we can gauge-fix the integral over random configurations to an integral over exactly closed polyhedral configurations (up to a measure zero set):
\beq
d_N[J]&=&
\f{1}{{J!(J+1)!}}
\int \prod_i\f{d^4 z_i}{\pi^2}\,
\delta\left(\sum_i|z_i\ra\la z_i|-\f12\sum_i\la z_i|z_i\ra\id\right)\,
\la J,\{z_i\}|J,\{z_i\}\ra
\int_{\SL(2,\C)}
e^{-\sum_i \la z_i|\Lambda^\dag\Lambda|z_i\ra }\nn\\
&=&
\f{1}{2^{2J}{J!(J+1)!}}
\int \prod_i\f{d^3 \vV_i}{4\pi V_i}\,
\mu\left(\{\vV_i\}\right)\,
\delta\left(\sum_i\vV_i\right)\,
(\sum V_i)^{2J}\,,
\eeq
where the measure $\mu(\{\vV_i\})$ reflects the volume of the gauge orbits under the $\SL(2,\C)$ action\footnotemark:
$$
\mu(\{\vV_i\})
\,=\,
\int_{\SL(2,\C)}
e^{-\f12(\sum_i V_i)\tr\Lambda^\dag\Lambda}
\quad=\,
\mu(\sum_i V_i)\,,
$$
where we have assumed that the $z_i$ and $\vV_i$ satisfy the closure constraint.
\footnotetext{
Writing explicitly the $\SL(2,\C)$-matrix $\Lambda$ in terms of its four matrix elements $a,b,c,d$ we can express the measure as a quasi-Gaussian integral in terms of $A=\f12\sum_i V_i$:
$$
\mu(\{\vV_i\})
\,=\,
\int d^2ad^2bd^2cd^2d\,
\delta^{(2)}(ad-bc-1)\,e^{-A(|a|^2+|b|^2+|c|^2+|d|^2)}
\,=\,
\int \f{dT}{2\pi}\int d^2ad^2bd^2cd^2d\,
e^{(iT-\eps)(ad-bc-1)}\,e^{-A(|a|^2+|b|^2+|c|^2+|d|^2)}\,,
$$
where $\eps$ is an arbitrary positive regulator (the integral does not depend on its value) and the integral over the $a,b,c,d$ gives simply the inverse square-root of the determinant of the Hessian. Then it still remains to compute the integral over $T$.
}
This measure factor is a function of only the total area $\sum_i V_i$, we will not compute it but we refer the interested reader to similar calculations in \cite{un2,conrady,tetra}.

Our purpose here is merely to point out the existence of such a formula giving the dimension of the intertwiner space as an integral over closed classical polyhedra but with a non-trivial measure. However, the integral over almost-closed configuration derived above is simply more convenient to handle for practical purposes.

\subsection{The Generating Functional as Integral over Fuzzy Polyhedra}

Interesting, we can also give an interesting geometrical interpretation for the generating functional $f_N[t]$ defined in \eqref{fN_def} as the sum over the total spin $J$. Starting with the expression \eqref{coh1} for the dimension of the intertwiner space in terms of $\SU(2)$ coherent states, the sum over $J$ allows to sum over all the individual spins $j_i$ and finally obtain a simpler formula:
\be
f_N[t]
\,=\,
\sum_{J\in\N}t^{2J}d_N[J]
\,=\,
\int_{\SU(2)}dg\,
\int \prod_i^N \f{d^4z_i}{\pi^2}\,
e^{-\la z_i|z_i\ra}e^{t\la z_i|g|z_i\ra}
\ee
Switching the integration from the spinors $z_i$ to the 3-vectors $\vV_i$ and parameterizing the group element $g=\cos\theta+i\sin\theta \hat{u}\dot\vsigma$ as before, we get:
\be
f_N[t]
\,=\,
\int \prod_i^N \f{e^{-V_i}d^3\vV_i}{4\pi V_i}\,
\int_0^\pi \f2\pi\sin^2\theta d\theta\int_{\cS_2}d^2\hat{u}\,
e^{\cos\theta\sum_i V_i}\,e^{i\sin\theta \hat{u}\cdot\sum_i\vV_i}\,.
\ee
Written as such, the integration over $\SU(2)$ truly resembles imposing the closure constraint $\sum_i\vV_i$, up to the fact that the Lagrange multiplier $\vec{p}\equiv \sin\theta \hat{u}$ has a norm bounded by $1$. We can thus interpret this formula as an integration over ``almost-closed polyhedra".

If we were to perform explicitly the integration over $\SU(2)$, we would actually get the expression derived using the $\U(N)$ coherent states:
\be
f_N[t]
\,=\,
\int \prod_i\f{e^{-V_i}d^3 V_i}{4\pi V_i}
\sum_J\f{t^{2J}}{2^{2J}{J!(J+1)!}}\left(\big(\sum_i V_i\big)^2-|\sum_i \vV_i|^2
\right)^J
\,=\,
2\int \prod_i\f{e^{-V_i}d^3 V_i}{4\pi V_i}\,
\f{I_1\left(t\,\sqrt{\sum_i V_i)^2-|\sum_i \vV_i|^2}\right)}{t\,\sqrt{\sum_i V_i)^2-|\sum_i \vV_i|^2}},\nn
\ee
where $I_1$ is a modified Bessel function of the first kind. However this expression does not make clear that the integration is peaked on closed polyhedral configurations.

%

\section{On the Role of the Number of Faces}
\label{discuss}

We have discussed a classical polymer model for the black hole horizon, modeling it as a classical polyhedron with $N$ faces. Quantizing this into a quantum polymer model leads to the description of quantum isolated horizon whose states are $N$-valent intertwiners in loop quantum gravity. We have computed the number of states for fixed area, giving a density of states $\rho_N[A]$ in the classical model and the number of intertwiners $d_N[J]$ where $A$ and $J$ are the total boundary area in the corresponding model. Then we would take the logarithm of this number of states and get the horizon entropy.
An obvious issue comparing the standard classical description of the black hole entropy is that we now have two numbers defining the properties of the horizon, the total area $A$ (or $J$) and the number of faces (or legs of the intertwiner) $N$. We thus need to comment on the role of $N$.

In the recent work \cite{alej}, it was proposed that $N$ be considered as some ``quantum hair", needed to define the properties of the black hole horizon at the quantum level and without any classical counterpart. Then, as was shown in \cite{un1}, if we want the entropy $\log \rho_N[A]$ or $\log d_N[J]$ to scale at leading order linearly with the total area, we need the number of faces $N$ to grow linearly with the total boundary $A$. In this case, the entropy/area ratio depends on the $N/A$ ratio. One can thus fine-tune this ratio in order to ensure the correct entropy/area ratio. Going further it was proposed in \cite{un1} that the ratio $N/A$ defines various regimes of the theory and should be selected by the actual quantum gravity dynamics or by the physical requirement of the surface to be a true black hole horizon. For instance, a black hole horizon could have a certain $N/A$ ratio while another type of surface has a different value of this ratio due the different quantum gravity dynamics in the neighborhood of the surface.

In such a scenario, the dynamics of the near-horizon geometry plays an essential role. Indeed the number of faces $N$ is actually in loop quantum gravity a property of the spin network states describing the exterior geometry and puncturing the horizon surface. It is the near-horizon geometry state and its dynamics that determines that the considered surface is actually a isolated horizon. The study of the near-horizon geometry plays actually an essential role in other approaches to the black hole entropy, such as the conformal field theory calculations (e.g. \cite{carlip}) or the more recent proposal \cite{recent}.
From this perspective, the near-horizon geometry state will determine what are $N$ and $A$. It is not unreasonable to expect that the specific shape and choice of the near-horizon wave-function leads to a relation between $N$ and $A$. Then in order to discuss the value of $N$ we would need to enlarge our polymer model for the horizon to take into account its coupling to the exterior geometry.

\smallskip

Another possibility is to consider that we have already taken into account and integrated over the dynamics of the exterior geometry and that it leads to a certain weight for the creation and annihilation of a face/puncture on the horizon. For this point of view, it is thus natural to sum over the number of faces $N$ together a certain statistical weight $\alpha^N$ reflecting the probability of the horizon of having $N$ faces. Such a weight implicitly assumes that the creation/annihilation of a single puncture is independent from the other faces. This might have to be revised once we have a decent model of the dynamics of the near-horizon geometry. Putting this aside, we are led to an entropy as we have defined in the present work $S[\alpha,A]=\log \sum_N \alpha^N \rho_N[A]$.

The parameter $\alpha$ is the chemical potential and controls the probability distribution of the number of faces.  We can actually look at this probability distribution in our polymer models and we easily find that for large areas it has approximatively a Gaussian shape peaked on a optimal value $N_{max}$ which depends on the chemical potential $\alpha$ and grows linearly with the horizon area $A$. This means that we actually recover effectively the same picture as in the previous scenario with the number of faces fluctuating and being determined dynamically to a most probable value growing with the area. Then the ratio $N_{max}/A$ depends on the chemical potential $\alpha$, which reflects the strength of the coupling between the horizon and the outside geometry.

Finally, in order to go further, we will need to go beyond the present kinematical description and look at the dynamics of the horizon and its coupling to the outside geometry state. Such a proposal was recently put forward in \cite{recent} where a quantum Hamiltonian for a quantum Rindler horizon in (covariant) loop quantum gravity was defined. This should not only lead to a better understanding of the role of the number of faces/punctures but will most importantly necessarily lead to a description of the thermal radiation and black hole evaporation process.

\section*{Conclusions}

Following the identification of black hole boundary states as $\SU(2)$ intertwiners and the the description of those intertwiners as quantized polyhedra, we introduced a classical polymer model, which defines the black hole entropy as a counting of classical polyhedra with fixed total area, and we have explained how its quantization leads to a quantum polymer model reproducing the usual LQG intertwiner entropy counting.
This shows that most of the features of the intertwiner counting are already present in a classical computation of the density of polyhedra at given area.

First, summing over the number of faces of the polyhedra, or equivalently the number of punctures on the horizon or the number of legs of the intertwiners, leads 
to a leading order entropy scaling with the total boundary area. The entropy-area ratio depends on the value of the chemical potential, which also controls the optimal number of faces in terms of the boundary area. We argue that the precise value of the chemical potential should be determined dynamically through the coupling of the polyhedra/intertwiner with the exterior geometry.

Second, we recover the expected $-\f32$ log-correction. We show that it comes directly from the closure constraint imposed on a set of $N$ vectors (in $\R^3$) for them to properly form a geometrical (convex) polyhedra (as normal vectors to the faces). This constraint is the classical counterpart of the requirement of $\SU(2)$ invariance at the quantum level, which defines intertwiner states. By a full and explicit analysis, we remove and relax the closure constraint and show that it affects the log-correction (but obviously not the leading  order). We do the same at the quantum level and show how removing or relaxing the requirement of $\SU(2)$ invariance changes the entropy at next-to-leading order.

Finally, we provide a semi-classical formula for the intertwiner counting as a density of almost-closed polyhedral configuration, which shows how to interpolate from the classical model to the fully quantum model.

As next steps, we see two directions. At the kinematical level, we could dig further than the mere entropy which counts the total number of states, but we should compute the correlations between the faces (or equivalently the punctures on the horizon or the legs of the intertwiners). This should allow us to probe the shape and structure of the horizon, thus probably understand how the angular momentum and higher multipolar moments (see \cite{multipolar} for the definition of multipole moments of isolated horizon) can be taken into account as suggested in \cite{eugenio}. But this could also give us some hints to the evaporation process, which seems to be related to the entanglement between the punctures on the horizon as suggested in \cite{danny}. Then we should go dynamical, model and study the near-horizon geometry and its coupling to the boundary horizon, in order to get a clear picture of the thermodynamical role of the entropy which we computed and the resulting Hawking radiation. This would truly validate, or invalidate, the physical interpretation of the mathematical polymer model for black hole horizons which we have discussed here.

\section*{Acknowledgments}

EL is partially supported by the ANR ``Programme Blanc" grants LQG-09.

Research at Perimeter Institute is supported by the Government of Canada through Industry Canada and by the Province of Ontario through the Ministry of Research and Innovation.

\appendix

\section{Weakening the Closure Constraints for the Classical Polymer Model}

\subsection{Ensemble Without the Closure Constraint}
\label{0closure}

Let us now investigate the effect of relaxing the closure constraint. Let us thus fix the total area and introduce the following density of states:
\be
\label{rho0_def}
\rho^0_N[A]
\,\equiv\,
\int\prod_i^N\f{e^{-\eps V_i}d^3\vV_i}{4\pi V_i}\,\delta\left(\sum_k V_k-2A\right)\,,
\ee
with the regulator $\eps$ as before.
We use the same technique as before. We take the Fourier transform of the $\delta$-distribution and swap the integrals:
\be
\rho^0_N[A]=
\int \f{dq}{2\pi}\,
e^{-2iqA}
I^0(q)^N
\qquad\textrm{with}\quad
I^0(q)\equiv\,
\int \f{d^3\vV}{4\pi V}\,e^{-\eps V}e^{iqV}
\,.
\ee
The integral kernel $I^0$ is simpler to compute than in the case with the closure constraints:
\be
I^0(q)=\int_0^{+\infty} VdV\,e^{-\eps V}\,e^{iqV}
=-\f{ 1}{(q+i\eps)^2}=\f{1}{(\epsilon-i q)^2}\,.
\ee
We can then perform the integral over $q$ and get:
\be
\rho_N[A]=
\int \f{dq}{2\pi}\,
e^{-2iqA}\f{1}{(\eps-iq)^{2N}}
\,=\,
e^{-2\eps A }\,\f{(2A)^{2N-1}}{(2N-1)!}.
\ee
The factor $e^{-2\eps A }$ becomes trivial when the regulator is sent to $\eps\arr 0^+$. The area dependence $A^{2N-1}$ is easily derived by doing the change of variable $\vV\arr \vV/A$ in the definition \eqref{rho0_def} of the density of state $\rho^0$.

As in the case of closed polyhedra, we define the generating functional $\rho^0[\alpha,A]$ by summing over the number of punctures $N$ with an arbitrary geometric weight $\alpha^N$:
\be
\rho^0[\alpha,A]\equiv\sum_{N\ge 1} \alpha^{N}\rho_N[A]
=\sum_{N=0}^{\infty}\alpha^{N+1}\f{(2A)^{2N+1}}{(2N+1)!}
=\sqrt{\alpha}\sinh( 2A\sqrt{\alpha}).
\ee
The entropy is defined as the logarithm of the density of states and it is straightforward to extract its asymptotical behavior at large area:
\be
S^0[\alpha,A]=\log \rho^0[\alpha,A]
=\log\sinh (2A\sqrt{\alpha})
\sim
2A\sqrt{\alpha}-\log 2\,.
\ee
A first remark is that we obviously check that the entropy without the closure constraint is larger than the entropy computed earlier with the closure constraints, $S^0[\alpha,A]>S[\alpha,A]$. Then the point which we wish to illustrate is that we do not have any log-correction term and that the asymptotical formula goes directly to the constant correction term. This is coherent with the interpretation that the log-correction is specifically due to the closure constraints.

\subsection{Weakening the Closure Constraint: $z$-Closure Model}
\label{zclosure}

Let us now introduce and analyze a model where we impose the closure constraint along a single axis, $\sum_i V_i^z=0$. This can be considered as the classical counterpart of the ``old" calculation for black hole entropy calculation in  Loop Quantum Gravity that is based on counting $\U(1)$ intertwiners \cite{old}.

We define the following density of states at fixed total area $A$ and satisfying the $z$-closure constraint:
\be
\rho^1_N[A]=\int\prod_i^N\f{e^{-\eps V_i}\,d^3\vV_i}{4\pi V_i}\,\delta\left(\sum_k V_k-2A\right)\,\delta\left(\sum_k V^z_k\right)
\ee
Taking the Fourier transform of the constraints it becomes
\be
\rho^1_N[A]=
\int \f{du}{(2\pi)}\,\f{dq}{2\pi}\,
e^{-2iqA}
I^1(q,u)^N
\qquad\textrm{with}\quad
I^1(q,u)\equiv\,
\int \f{d^3\vV}{4\pi V}\,e^{-\eps V}e^{iqV}e^{iuV^z}\,. \label{rho-1}
\ee
This new integral $I^1(q,u)$ is actually equal to the integral $I(q,\vu)$ with the full closure constraints:
\be
I^1(q,u)=\int_0^{+\infty} V dV\,e^{-\eps V}e^{iqV}\,\f{\sin uV}{uV}
=\f{1}{u^2-(q+i\eps)^2}\,.
\ee
The difference between the expressions \eqref{rho-1} and \eqref{rho-full} is that we are now integrating over a single variable $u\in\R$ instead of a 3-vector $\vu\in\R^3$. This gives:
\beq
\rho^1_N[A]&=&
\int_{\R}\f{dq}{2\pi}\,e^{-2iqA}\int_{-\infty}^{+\infty} \f{ du}{2\pi}\,
\f{1}{(u^2-(q+i\eps)^2)^N}\nn\\
&=&
\int_{\R}\f{dq}{2\pi}\,e^{-2iqA}\,
\f{(\eps-iq)^{1-2N}}{2^{2N-1}}\bin{N-1}{2N-2}\nn\\
&=&
\f12\,e^{-2A\eps}\,\f{A^{2N-2}}{(N-1)!^2}.
\eeq
As before, the factor $e^{-2A\eps}$ vanishes as the regulator $\eps$ is sent to 0. The power of the area, $A^{2N-2}$, can be again simply derived by performing the change of variable $\vV_i\arr \vV_i/A$ in the initial definition of the density $\rho^1_N[A]$. The difference with the case with the full closure constraints, besides the slight shift in the power of $A$, is the change of the $(N-2)!$-factor with a $(N-1)!$-factor . This little difference will be responsible for a change in the log-correction.

Indeed, we now define the generating functional by summing the number of faces $N$ with the geometric weight $\alpha^N$:
\be
\rho^1[\alpha,A]
\equiv
\rho^1_1[A]+\alpha \rho^1_2[A]+\alpha^2 \rho^1_3[A]+\dots
\,=\,
\sum_{N\ge 1}\alpha^{N-1}\rho^1_N[A]
\,=\,
\f12\sum_{N=0}^\infty \alpha^{N}\f{A^{2N}}{N!^2}
\,=\,
\f12I_0(2A\sqrt{\alpha})\,,
\ee
where $I_0(x)$ is the modified Bessel function of zeroth order.
We take the logarithm of this density of states to define the entropy. It is then straightforward to derive its asymptotic expansion:
\be
S^1[\alpha,A]\equiv\log\rho^1[\alpha,A]
\sim
2A\sqrt{\alpha}-\f12\log A-\f14\log\alpha-\log4-\f12\log\pi+\dots
\ee
We find the same leading order as with and without the full closure constraints. The main difference is with the next-to-leading order, where the log-correction now comes with a $-\f12$ factor.

\section{Weakening the Closure Constraints at the Quantum Level}
\label{relax}

\subsection{Without the Closure Constraint}

Let us start with the quantum polymer model and remove the closure constraints as in the classical case (section \ref{0closure}) and see what happens. Now we do not require the invariance under $\SU(2)$ and our Hilbert space for $N$ faces and fixed total area $J$ is:
\be
H^N_J=\bigoplus_{\sum j_i =J} V^{j_1}\otimes..\otimes V^{j_N}\,.
\ee
Its dimension is obviously:
\be
d^0_N[J]=\sum_{\sum j_i =J}\prod_i^N (2j_i+1)
=\sum_{\sum j_i =J}\prod_i^N \chi^{j_i}(\id)\,,
\ee
which can be interpreted as the dimension for the intertwiner space, but trading the group averaging integral over $\SU(2)$ for a localized evaluation on the identity $g=\id$.
The generating functionals  can be computed similarly. We thus define and compute the various generating functionals summing alternatively or simultaneously over the number of faces $N$ and the total area $J$:
\be
d^0[\alpha,J]\,\equiv\,
\sum_N\alpha^N d^0_N[J]\,,
\ee
\be
f^0_N[t]
\,\equiv\,
\sum_{J\in\N} t^{2J} d^0_N[J]
\,=\,
\left(
\sum_{j\in\N/2} t^{2j}(2j+1)
\right)^N
\,=\,
\left(
\f1{(1-t)^2}
\right)^N\,,
\ee
\be
F^0[\alpha,t]
\,\equiv\,
\sum_{N\in\N} \alpha^{N}f^0_N[t]
\,=\,
\sum_{J} t^{2J}d^0[\alpha,J]
\,=\,
\sum_{N} \alpha^{N}\f1{(1-t)^{2N}}
\,=\,
\f{(1-t)^2}{t^2-2t+1-\alpha}\,.
\ee
The lowest pole in $t$ in terms of the variable $u$ is easily identified, $t_c=1-\sqrt{\alpha}$, and controls the asymptotics of $d^0[\alpha,J]$. Thus we get for large area $J$:
\be
\cS^0[\alpha,J]\,\equiv\,\log d^0[\alpha,J]
\sim
-2J\log(1-\sqrt{\alpha})+\dots
\ee
This reproduces the exact same leading order as in the intertwiner counting with the $\SU(2)$-invariance. Thus this leading order, which gives the area-entropy law, is not related in any way to the requirement of $\SU(2)$-gauge invariance but simply to the straightforward sum over all possible spins on the $N$ faces on the horizon.

One can go further and provide exact formulas for the dimensions $d^0_N[J]$:
\be
d^0_N[J]=\mat{c}{2N+2J-1 \\ 2J},\qquad
d^0[\alpha, J]=\sum_{N\in\N}\alpha^N\,\mat{c}{2N+2J-1 \\ 2J}\,.
\ee
From these expressions, it is straightforward to check that the log-correction to the entropy vanishes for large $J$.

\subsection{Weakening the Closure Constraint: Quantum $z$-Closure}

Let us now introduce the quantum model corresponding to a single closure constraint along the $z$ axis. This amounts to drop the $\SU(2)$-gauge invariance and replace it by the requirement of gauge invariance under $\U(1)$. This actually corresponds to the initial black hole entropy counting in Loop Quantum Gravity based on the identification of a $\U(1)$ Chern-Simons theory living of the black hole horizon \cite{old}.

The number of states for fixed number of faces $N$ and total area $J$ is now derived through a group averaging over $\U(1)$:
\be
\label{dN1_def}
d_N^1[J]
\,=\,
\int_{\U(1)} dh\,
\sum_{\sum_i^N j_i=J} \prod_i^N \chi_{j_i}(h)\,,
\ee
where $\U(1)$ group elements are generated by the $\su(2)$ generator $J_z$, more explicitly $h=e^{i \phi J_z}$. We further define the sum $d^1[\alpha,J]=\sum_{N\in\N} \alpha^{N}d_N^1[J]$ over the number of faces $N$. As before, we define the generating functionals:
\be
f_N^1[t]
\,\equiv\,
\sum_{J\in\N} t^{2J} d_N^1[J]
\,=\,
\int_{\U(1))}dh\,\left(
\sum_{j\in\N/2} t^{2j}\chi^{j}(h)
\right)^N\,,
\ee
\be
F^1[\alpha,t]
=\sum_J t^{2J} \sum_{N\in\N} \alpha^{N}d_N^1[J]
=\sum_{N\in\N} \alpha^{N}f_N^1[t]
=\sum_J t^{2J} d^1[\alpha,J]
\,.
\ee
%
%
Writing the integrals explicitly in terms of the class angle $\phi$ of the group element, we get:
\be
f_N^1[t]
\,=\,
\int_0^{2\pi} \f{d\phi}{2\pi}\,\left(
\sum_{j\in\N/2} t^{2j}\f{\sin(2j+1)\phi}{\sin\phi}
\right)^N
\,=\,
\int_0^{2\pi} \f{d\phi}{2\pi}\,
\f{1}{(t^2-2t\cos\phi+1)^N}\,,
\ee
\be
F^1[\alpha,t]=
\int_0^{2\pi} \f{d\phi}{2\pi}\,
\f{(t^2-2t\cos\phi+1)}{(t^2-2t\cos\phi+1)-\alpha}\,.
\ee
The main difference with the $\SU(2)$-invariant calculations \eqref{fN_int} and \eqref{F_int} is the measure term, where the $\sin^2\phi$ factor now drops out.
This integral can be computed explicitly from $\alpha,t\sim 0$:
\be
F^1[\alpha,t]
\,=\,
1+\f\alpha{\sqrt{(\alpha-(1+t)^2)(\alpha-(1-t)^2)}}
\,=\,
1+\f\alpha{\sqrt{t^4-2(1+\alpha)t^2+(1-\alpha)^2}}\,.
\ee
As before the lowest pole in $t$ at fixed $\alpha\in]0,1[$ is easily identified at $t_c=1-\sqrt{\alpha}$, which leads to the same asymptotics at leading order for the entropy:
\be
\cS^1[\alpha]
\,\equiv\,
\log d_N^1[\alpha,J]
\,=\,
\log\sum_{N} \alpha^{N}d_N^1[J]
\,\underset{J\gg1}\sim\,
-2J\log (1-\sqrt{\alpha})+\dots
\ee
One can go further and (easily) identify a first order differential equation satisfied by the generating functional $F^1[\alpha,t]$. Indeed, calling $T=t^2$, we have:
\be
(T^2-2(1+\alpha)T+(1-\alpha)^2)\,\pp_T F^1
+(T-(1+\alpha))(F-1)=0\,.
\ee
This naturally induces a recursion relation on the coefficient of the series expansion $F^1=\sum_J T^{J} d^1[\alpha,J]$~:
\be
\forall J\ge 2,\qquad
(1-\alpha)^2(J+1) d^1[\alpha,J+1]
-(1+\alpha)(2J+1) d^1[\alpha,J]
+j d^1[\alpha,J-1]=0\,.
\ee
Inserting the ansatz $d^1[\alpha,J]\sim A^J/J^\sigma$ for large area $J$ in this recursion relation, we get equations for the parameters $A$ and $\sigma$ at leading and next-to-leading order in $J$:
$$
(1-\alpha)^2A-2(1+\alpha)+\f1A=0,\qquad
(1-\sigma)(1-\alpha)^2A-(1+\alpha)+\sigma\f1A=0\,.
$$
These equations are very simple to solve and give:
\be
A=\f1{(1-\sqrt\alpha)^2},\qquad
\sigma=\f12\,,
\ee
which gives the following asymptotics for the entropy:
\be
\cS^1[\alpha]
\,=\,
\log d_N^1[\alpha,J]
\,\underset{J\gg1}\sim\,
-2J\log (1-\sqrt{\alpha})-\f12\log J,
\ee
with the expected factor $-\f12$ for the log-correction.


\section{Bianchi's Semi-Quantized Model or Bohr-Sommerfeld Approximation}
\label{BS}

As we see in Sec.~III, the analysis for the classical polymer model for the density of states $\rho_N[A]$ corresponds to the first term in the expansion of the exact quantum result $\dim_J[N]$ for large  area $J$ at fixed number of faces $N$. An intermediate model was proposed in \cite{eugenio} where the area of faces are quantized while the direction of their normal vector remains continuous and arbitrary on the 2-sphere. This half-quantized model, where the normal vectors do not become true quantum vectors but nevertheless have a discrete norm, can be seen as a Bohr-Sommerfeld approximation of the fully quantized model.

This model is not particularly relevant to the present discussions, but we include its definition and a quick analysis for the sake of completeness.
Hence we assume that areas of the individual faces are quantized $V_i=a m_i$ with $m_i\in\N$ and a fixed area unit $a$ (with dimensions restored) being of the order of the Planck area. But the directions of the individual vectors remain unrestricted. The total area is discrete in $a$-unit, now  $2A=2aJ$.
Then in this ``Bohr-Sommerfeld" approximation the density of states is given  as:
\be
\rho^{BS}_N[M]=4\pi\sum_{\{m_i\}}\int\!\left(\prod_{i=1}^N m_i a^2 \f{d^2\hat{V}_i}{4\pi}\right)\frac{1}{a^4}\,\delta\!\left(\sum_k m_k-2J\right)\delta^{(3)}\!\left(\sum_k m_k\hat{V}_k\right), \label{bs-def}
\ee
where a $\delta$-function on integers is the Kronecker $\delta$. The the normalization is again chosen to enforce $\rho^{BS}_2[J]=1$. Proceeding analogously to the derivation of Sec.~\ref{computerho} we write
\be
\rho^{BS}_N[J]=8\pi\int_{-\frac{\pi}{a}}^{\frac{\pi}{a}}\!\frac{dq}{2\pi}\int\!\f{d^3\vu}{(2\pi)^3}
\sum_{\{m_i\}}\int\!\left(\prod_{i=1}^N m_i a^2 e^{-a m_i \epsilon}d^2\hat{V}_i\right)e^{i a\vu\cdot\sum_k m_k\hat{V}_k}
e^{i a q(\sum_k m_k V_k-2 J)}. \label{bs-fou}
\ee
This leads to:
\be
\rho^{BS}_N[J]=8\pi\int_{-\frac{\pi}{a}}^{\frac{\pi}{a}}\!\frac{dq}{2\pi}\int\!\f{d^3\vu}{(2\pi)^3}K(q,\vu)^N e^{-2 i a q J}\,,
\ee
where the new kernel $K(q,\vu)$ results from the integration over the normal vectors:
\be
K(q,\vu)
\,\equiv\,
\sum_{m=0}^\infty m a^2e^{-\epsilon a m}e^{i q a m}\int\! \f{d^2\hat{V}}{4\pi}e^{i a \hat{V}\cdot\vu}
\,=\,
\f{\sin au}{au}\f{a^2}{2(\cosh(a(\epsilon-i q))-\cos a u)}.
\ee
We haven't found a systematic way to perform the exact integration over the Lagrange multipliers $q$ and $\vu$ for all $N$ and $J$ so we are unable to provide an explicit closed analytic formula for the density of states in this model. We can nevertheless investigate the leading order effect of the area quantization in the $N\ll J\rightarrow\infty$ regime where we keep $N$ fix while considering large total areas $J$.

Taking the $N$-power of ${\sin au}/{au}$ ensures that the dominant contribution to $\rho^{BS}_N[M]$ comes from $a\sqrt{N} u\lesssim 1$, allowing to disregard the poles at $u=q+2\pi n/a +i \epsilon$ for $n>1$.
In this regime we approximate the new kernel as:
\be
K(q,\vu)=\f{1}{u^2-(q+i\epsilon)^2}-\f{\,a^2}{12}+{\cal{O}}(a^4)
\approx I(q,\vu)-a^2/12.
\ee
Then the BS density of state with the first non-trivial correction in the area unit $a$  is given  by:
\be
\rho^{BS}_N[J]\approx
8\pi\int_{-\frac{\pi}{a}}^{\frac{\pi}{a}}\!\frac{dq}{2\pi}\int\!\f{d^3\vu}{(2\pi)^3}
e^{-2 i a q J}
\left(\f{1}{(u^2-(q+i\eps)^2)^N}-
\f{N}{12}\f{a^2}{(u^2-(q+i\eps)^2)^{N-1}}\right).
\ee
Performing the integration (and taking the limit $\pi/a\rightarrow\infty)$ we find
\be
\rho^{BS}_N[A=2aJ]\approx \rho_N[A=2aJ]\left(1-\frac{N^3}{J^2}\right)+\ldots.
\ee
This expression obviously breaks down when $N\sim J^{2/3}$, but we insist that this approximation holes for fixed $N$ while $J$ is sent to infinity.
So this quantized area model obviously fits with the classical model at leading order, but its first correction in $a$ (beyond the leading order in $N\ll J$) unfortunately  does not fit with the exact  quantum expression (Sec.~III), and thus this BS model will not be particularly relevant to our present work.



\begin{thebibliography}{99}

\bibitem{isolated}
A. Ashtekar, C. Beetle and S. Fairhurst,
{\it Mechanics of Isolated Horizons},
Class.Quant.Grav. 17 (2000) 253-298 [arXiv:gr-qc/9907068];\\
A. Ashtekar and B. Krishnan,
{\it Isolated and dynamical horizons and their applications},
LivingRev.Rel.7 (2004) 10 [arXiv:gr-qc/0407042]


\bibitem{old}
A. Ashtekar, J. Baez and K. Krasnov,
{\it Quantum Geometry of Isolated Horizons and Black Hole Entropy},
Adv.Theor.Math.Phys. 4 (2000) 1-94  [arXiv:gr-qc/0005126];\\
A. Ashtekar, J. Baez, A. Corichi and K. Krasnov,
{\it Quantum Geometry and Black Hole Entropy},
Phys.Rev.Lett. 80 (1998) 904-907 [arXiv:gr-qc/9710007]

\bibitem{alej0}
J. Engle, K. Noui and A. Perez,
{\it Black hole entropy and SU(2) Chern-Simons theory},
arXiv:0905.3168;\\
J. Engle, K. Noui, A. Perez and D. Pranzetti,
{\it Black hole entropy from an SU(2)-invariant formulation of Type I isolated horizons},
arXiv:1006.0634;\\
A. Perez and D. Pranzetti,
{\it Static isolated horizons: SU(2) invariant phase space, quantization, and black hole entropy},
arXiv:1011.2961

\bibitem{kaul}
R. Kaul,
{\it Entropy of Quantum Black Holes},
SIGMA 8 (2012) 005 [arXiv:1201.6102]

\bibitem{kaul2}
R.K. Kaul and P. Majumdar,
{\it Quantum Black Hole Entropy},
Phys.Lett. B439 (1998) 267-270 [arXiv:gr-qc/9801080];\\
R.K. Kaul and P. Majumdar,
{\it Black Hole Entropy from Spin One Punctures},
Phys.Rev. D68 (2003) 024001  [arXiv:gr-qc/0301128]

\bibitem{danny}
E.R. Livine and D. R. Terno,
{\it Quantum Black Holes: Entropy and Entanglement on the Horizon},
Nucl.Phys.B741 (2006) 131-161 [arXiv:gr-qc/0508085]

\bibitem{spanish}
I. Agullo, J.F. Barbero G., E.F. Borja, J. Diaz-Polo and E.J.S. Villase\~nor,
{\it The combinatorics of the SU(2) black hole entropy in loop quantum gravity},
Phys.Rev.D80 (2009) 084006 [arXiv:0906.4529];\\
I. Agullo, J.F. Barbero G., E.F. Borja, J. Diaz-Polo and E.J.S. Villase\~nor,
{\it Detailed black hole state counting in loop quantum gravity},
Phys.Rev.D82 (2010) 084029 [arXiv:1101.3660]

\bibitem{un1}
L. Freidel and E.R. Livine,
{\it The Fine Structure of SU(2) Intertwiners from U(N) Representations},
J.Math.Phys. 51 (2010) 082502 [arXiv:0911.3553]

\bibitem{alej}
A. Ghosh and A. Perez,
{\it Black hole entropy and isolated horizons thermodynamics},
arXiv:1107.1320

\bibitem{mitra}
A. Ghosh and P. Mitra,
{\it Black hole state counting in loop quantum gravity},
arXiv:1105.6034;\\
P. Mitra,
{\it  Area law for black hole entropy in the SU(2) quantum geometry approach},
Phys. Rev. D85 (2012) 104025 [arXiv:1107.4605]

\bibitem{un0}
F. Girelli and E.R. Livine,
{\it Reconstructing Quantum Geometry from Quantum Information: Spin Networks as Harmonic Oscillators},
Class.Quant.Grav. 22 (2005) 3295-3314 [arXiv:gr-qc/0501075]

\bibitem{coherent}
E.R. Livine and S. Speziale,
{\it A new spinfoam vertex for quantum gravity},
Phys.Rev.D76 (2007) 084028 [arXiv:0705.0674]

\bibitem{un2}
L. Freidel and E.R. Livine,
{\it U(N) Coherent States for Loop Quantum Gravity},
J.Math.Phys.52 (2011) 052502 [arXiv:1005.2090]

\bibitem{polyhedron}
E. Bianchi, P. Dona and S. Speziale,
{\it Polyhedra in loop quantum gravity},
Phys.Rev.D83 (2011) 044035 [arXiv:1009.3402]

\bibitem{eugenio}
E. Bianchi,
{\it Black Hole Entropy, Loop Gravity, and Polymer Physics},
Class.Quant.Grav.28 (2011) 114006 [arXiv:1011.5628]

\bibitem{spinor}
E.F. Borja, L. Freidel, I. Garay and E.R. Livine,
{\it U(N) tools for Loop Quantum Gravity: The Return of the Spinor},
Class.Quant.Grav.28 (2011) 055005 [arXiv:1010.5451]

\bibitem{un4}
M. Dupuis and E.R. Livine,
{\it Holomorphic Simplicity Constraints for 4d Spinfoam Models},
Class.Quant.Grav. 28 (2011) 215022 [arXiv:1104.3683]

\bibitem{merce}
E.R. Livine and M. Martin-Benito,
{\it Classical Setting and Effective Dynamics for Spinfoam Cosmology},
arXiv:1111.2867

\bibitem{spinor2}
E.R. Livine and J. Tambornino,
{\it Spinor Representation for Loop Quantum Gravity},
to appear in JMP (2012) [arXiv:1105.3385]

\bibitem{spinor3}
E.R. Livine and J. Tambornino,
{\it Loop gravity in terms of spinors},
Proceedings of Loops '11 (Madrid), to appear in Journal of Physics: Conference Series (JPCS) 2012 [arXiv:1109.3572]

\bibitem{bulk}
E.R. Livine and D.R. Terno,
{\it Bulk Entropy in Loop Quantum Gravity},
Nucl.Phys.B794 (2008) 138-153 [arXiv:0706.0985]


\bibitem{older}
A. Ghosh and P. Mitra,
{\it Counting of Black Hole Microstates},
Indian J.Phys. 80 (2006) 867 [arXiv:gr-qc/0603029]

\bibitem{area_krasnov}
K. Krasnov,
{\it The Area Spectrum in Quantum Gravity},
Class.Quant.Grav.15 (1998) L47-L53 [arXiv:gr-qc/9803074]

\bibitem{eprl-fk}
J. Engle, E. Livine, R. Pereira and C. Rovelli,
{\it LQG vertex with finite Immirzi parameter},
Nucl.Phys.B799 (2008) 136-149 [arXiv:0711.0146];\\
L. Freidel and K. Krasnov,
{\it A New Spin Foam Model for 4d Gravity},
Class.Quant.Grav.25 (2008) 125018 [arXiv:0708.1595]


\bibitem{conrady}
F. Conrady and L. Freidel,
{\it Quantum geometry from phase space reduction},
J.Math.Phys.50 (2009) 123510 [arXiv:0902.0351]

\bibitem{tetra}
L. Freidel, K. Krasnov and E.R. Livine,
{\it Holomorphic Factorization for a Quantum Tetrahedron},
Commun.Math.Phys.297 (2010) 45-93 [arXiv:0905.3627]

\bibitem{carlip}
S. Carlip,
{\it Horizon constraints and black hole entropy},
arXiv:gr-qc/0508071;\\
S. Carlip,
{\it Horizons, Constraints, and Black Hole Entropy },
Int.J.Theor.Phys.46 (2007) 2192-2203 [arXiv:gr-qc/0601041]

\bibitem{recent}
E. Frodden, A. Ghosh and A. Perez,
{\it A local first law for black hole thermodynamics},
arXiv:1110.4055

\bibitem{multipolar}
A.Ashtekar, J. Engle, T. Pawlowski and C. Van Den Broeck,
{\it Multipole Moments of Isolated Horizons},
Class.Quant.Grav. 21 (2004) 2549-2570 [arXiv:gr-qc/0401114]


\end{thebibliography}
\end{document}